\documentclass[twocolumn,times]{aastex63}

\newcommand\Cii{\hbox{[C\,{\sc ii}]}}
\newcommand\Hi{\hbox{H\,{\sc i}}}

\newcommand\comapeor{COMAP-\emph{EoR}}
\newcommand\comapera{COMAP-\emph{ERA}}
\newcommand{\avg}[1]{\left\langle{#1}\right\rangle}

\shorttitle{COMAP Early Science: I. Overview}
\shortauthors{Cleary et al.}
\graphicspath{{./}{figures/}}

\NewPageAfterKeywords

\begin{document}

\title{COMAP Early Science: I.\ Overview}

\correspondingauthor{Kieran A.\ Cleary}
\email{kcleary@astro.caltech.edu}

\author[0000-0002-8214-8265]{Kieran A.~Cleary}
\affiliation{California Institute of Technology, 1200 E. California Blvd., Pasadena, CA 91125, USA}


\author{Jowita Borowska}
\affiliation{Institute of Theoretical Astrophysics, University of Oslo, P.O. Box 1029 Blindern, N-0315 Oslo, Norway}

\author[0000-0001-8382-5275]{Patrick C.~Breysse}
\affiliation{Center for Cosmology and Particle Physics, Department of Physics, New York University, 726 Broadway, New York, NY, 10003, USA}

\author{Morgan Catha}
\affil{Owens Valley Radio Observatory, California Institute of Technology, Big Pine, CA 93513, USA}

\author[0000-0003-2618-6504]{Dongwoo T.~Chung}
\affil{Canadian Institute for Theoretical Astrophysics, University of Toronto, 60 St. George Street, Toronto, ON M5S 3H8, Canada}
\affil{Dunlap Institute for Astronomy and Astrophysics, University of Toronto, 50 St. George Street, Toronto, ON M5S 3H4, Canada}

\author[0000-0003-2358-9949]{Sarah E.~Church}
\affiliation{Kavli Institute for Particle Astrophysics and Cosmology \& Physics Department, Stanford University, Stanford, CA 94305, US}

\author{Clive Dickinson}
\affil{Jodrell Bank Centre for Astrophysics, Department of Physics and Astronomy, The University of Manchester, Oxford Road, Manchester, M13 9PL, U.K.}

\author[0000-0003-2332-5281]{Hans Kristian Eriksen}
\affiliation{Institute of Theoretical Astrophysics, University of Oslo, P.O. Box 1029 Blindern, N-0315 Oslo, Norway}

\author[0000-0001-8896-3159]{Marie Kristine Foss}
\affiliation{Institute of Theoretical Astrophysics, University of Oslo, P.O. Box 1029 Blindern, N-0315 Oslo, Norway}

\author{Joshua Ott Gundersen}
\affil{Department of Physics, University of Miami, 1320 Campo Sano Avenue, Coral Gables, FL 33146, USA}

\author[0000-0001-7911-5553]{Stuart E.~Harper}
\affil{Jodrell Bank Centre for Astrophysics, Department of Physics and Astronomy, The University of Manchester, Oxford Road, Manchester, M13 9PL, U.K.}

\author[0000-0001-6159-9174]{Andrew I.~Harris}
\affil{Department of Astronomy, University of Maryland, College Park, MD 20742}

\author{Richard Hobbs}
\affil{Owens Valley Radio Observatory, California Institute of Technology, Big Pine, CA 93513, USA}

\author[0000-0003-3420-7766]{H\aa vard T.~Ihle}
\affiliation{Institute of Theoretical Astrophysics, University of Oslo, P.O. Box 1029 Blindern, N-0315 Oslo, Norway}

\author[0000-0002-4274-9373]{Junhan Kim}
\affiliation{California Institute of Technology, 1200 E. California Blvd., Pasadena, CA 91125, USA}

\author{Jonathon Kocz}
\affil{California Institute of Technology, 1200 E. California Blvd., Pasadena, CA 91125, USA}
\affil{Department of Astronomy, University of California, Berkeley, CA, 94720, USA}

\author[0000-0002-5959-1285]{James W.~Lamb}
\affil{Owens Valley Radio Observatory, California Institute of Technology, Big Pine, CA 93513, USA}

\author{Jonas G.\ S.\ Lunde}
\affiliation{Institute of Theoretical Astrophysics, University of Oslo, P.O. Box 1029 Blindern, N-0315 Oslo, Norway}

\author[0000-0002-8800-5740]{Hamsa Padmanabhan}
\affiliation{D\'epartement de Physique Théorique, Universite de Genève, 24 Quai Ernest-Ansermet, CH-1211 Genève 4, Switzerland}

\author[0000-0001-5213-6231]{Timothy J.~Pearson}
\affiliation{California Institute of Technology, 1200 E. California Blvd., Pasadena, CA 91125, USA}

\author[0000-0001-7612-2379]{Liju Philip}
\affil{Jet Propulsion Laboratory, California Institute of Technology, 4800 Oak Grove Drive, Pasadena, CA 91109, USA}

\author{Travis W.~Powell}
\affil{Owens Valley Radio Observatory, California Institute of Technology, Big Pine, CA 93513, USA}

\author{Maren Rasmussen}
\affiliation{Institute of Theoretical Astrophysics, University of Oslo, P.O. Box 1029 Blindern, N-0315 Oslo, Norway}

\author[0000-0001-9152-961X]{Anthony C.S.~Readhead}
\affiliation{California Institute of Technology, 1200 E. California Blvd., Pasadena, CA 91125, USA}

\author[0000-0002-1667-3897]{Thomas J. Rennie}
\affil{Jodrell Bank Centre for Astrophysics, Department of Physics and Astronomy, The University of Manchester, Oxford Road, Manchester, M13 9PL, U.K.}

\author[0000-0003-0209-4816]{Marta B.~Silva}
\affiliation{Institute of Theoretical Astrophysics, University of Oslo, P.O. Box 1029 Blindern, N-0315 Oslo, Norway}

\author[0000-0001-5301-1377]{Nils-Ole Stutzer}
\affiliation{Institute of Theoretical Astrophysics, University of Oslo, P.O. Box 1029 Blindern, N-0315 Oslo, Norway}

\author[0000-0001-8526-3464]{Bade D.~Uzgil}
\affiliation{California Institute of Technology, 1200 E. California Blvd., Pasadena, CA 91125, USA}

\author[0000-0002-5437-6121]{Duncan J.~Watts}
\affiliation{Institute of Theoretical Astrophysics, University of Oslo, P.O. Box 1029 Blindern, N-0315 Oslo, Norway}

\author[0000-0003-3821-7275]{Ingunn Kathrine Wehus}
\affiliation{Institute of Theoretical Astrophysics, University of Oslo, P.O. Box 1029 Blindern, N-0315 Oslo, Norway}

\author{David P.~Woody}
\affil{Owens Valley Radio Observatory, California Institute of Technology, Big Pine, CA 93513, USA}


\author{Lilian Basoalto}
\affil{CePIA, Departamento de Astronomía, Universidad de Concepción, Chile}

\author[0000-0003-2358-9949]{J.~Richard Bond}
\affiliation{Canadian Institute for Theoretical Astrophysics, University of Toronto, 60 St. George Street, Toronto, ON M5S 3H8, Canada}

\author[0000-0002-5223-8315]{Delaney A.~Dunne}
\affiliation{California Institute of Technology, 1200 E. California Blvd., Pasadena, CA 91125, USA}

\author{Todd Gaier}
\affiliation{Jet Propulsion Laboratory, California Institute of Technology, 4800 Oak Grove Drive, Pasadena, CA 91109, USA}

\author[0000-0001-7449-4638]{Brandon Hensley}
\affiliation{Department of Astrophysical Sciences, Princeton University, Princeton, NJ 08544, USA}

\author[0000-0001-5211-1958]{Laura C.~Keating}
\affiliation{Leibniz-Institut f{\"u}r Astrophysik Potsdam (AIP), An der Sternwarte 16, D-14482 Potsdam, Germany}

\author{Charles R.~Lawrence}
\affiliation{Jet Propulsion Laboratory, California Institute of Technology, 4800 Oak Grove Drive, Pasadena, CA 91109, USA}

\author{Norman Murray}
\affiliation{Canadian Institute for Theoretical Astrophysics, University of Toronto, 60 St. George Street, Toronto, ON M5S 3H8, Canada}

\author[0000-0001-5704-271X]{Rodrigo Reeves}
\affil{CePIA, Departamento de Astronomía, Universidad de Concepción, Chile}

\author[0000-0003-0545-4872]{Marco P.~Viero}
\affiliation{California Institute of Technology, 1200 E. California Blvd., Pasadena, CA 91125, USA}

\author[0000-0003-2229-011X]{Risa H.~Wechsler}
\affiliation{Kavli Institute for Particle Astrophysics and Cosmology \& Physics Department, Stanford University, Stanford, CA 94305, US}


\collaboration{42}{(COMAP Collaboration)}



\begin{abstract}
The CO Mapping Array Project (COMAP) aims to use line intensity mapping of carbon monoxide (CO) to trace the distribution and global properties of galaxies over cosmic time, back to the Epoch of Reionization (EoR). To validate the technologies and techniques needed for this goal, a Pathfinder instrument has been constructed and fielded. Sensitive to CO(1--0) emission from $z=2.4$--$3.4$ and a fainter contribution from CO(2--1) at $z=6$--8, the Pathfinder is surveying $12$\,deg$^2$ in a 5-year observing campaign to detect the CO signal from $z\sim3$. Using data from the first 13 months of observing, we estimate $P_\mathrm{CO}(k)  = -2.7 \pm 1.7 \times 10^4\mu\textrm{K}^2\,\mathrm{Mpc}^3$ on scales $k=0.051-0.62 \,\mathrm{Mpc}^{-1}$ --- the first direct 3D constraint on the clustering component of the CO(1--0) power spectrum. Based on these observations alone, we obtain a constraint on the amplitude of the clustering component (the squared mean CO line temperature–bias product) of $\avg{Tb}^2<49$\,$\mu$K$^2$ --- nearly an order-of-magnitude improvement on the previous best measurement. These constraints allow us to rule out two models from the literature. We forecast a detection of the power spectrum after 5 years with signal-to-noise ratio (S/N) 9--17. Cross-correlation with an overlapping galaxy survey will yield a detection of the CO--galaxy power spectrum with S/N of 19. We are also conducting a 30\,GHz survey of the Galactic plane and present a preliminary map. Looking to the future of COMAP, we examine the prospects for future phases of the experiment to detect and characterize the CO signal from the EoR.

\end{abstract}




\section{Introduction} \label{sec:intro}
\begin{figure*}[t!]
\centering
\includegraphics[width=1.0\textwidth]{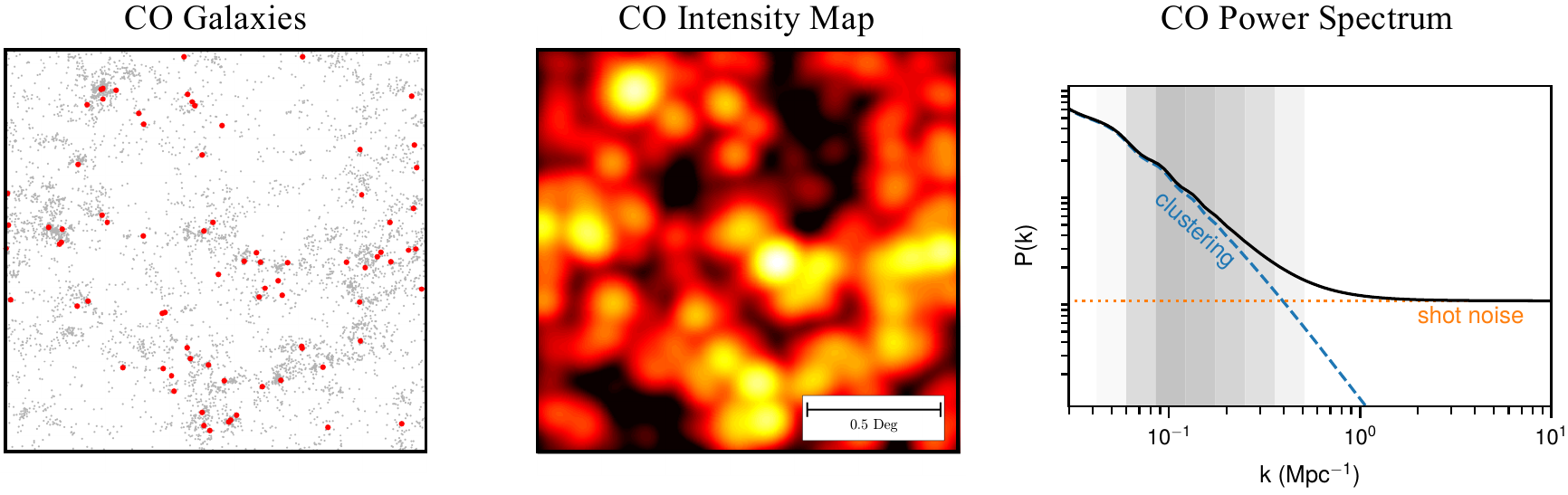}
\caption{{\bf\em Left: }A simulated 2.5 deg$^{2}$ field showing galaxy positions in gray (adapted from \citealt{kovetz_etal_17}). {\bf\em Center: }Simulated CO intensity map of the same field in a slice of 40 MHz bandwidth, corresponding to a redshift interval $\Delta z =0.004$. The VLA would take about 4,500 hours to cover the same area, but would detect just 1\% of the galaxies (shown in red on the left). COMAP, on the other hand, is sensitive to the aggregate emission from all galaxies in the line of sight, including those too faint to detect individually. {\bf\em Right: }A representative power spectrum for the intensity map shown in the center panel. The spectrum is composed of two components: one from the clustering of galaxies on large scales and a second that arises from the scale-independent shot noise, which dominates on small scales. The shaded region indicates schematically the scales to which the Pathfinder is most sensitive.
}\label{fig:flowchart}
\end{figure*}
Understanding the origin and evolution of the first stars and galaxies, from Cosmic Dawn to the present day, is a major challenge for astrophysics and cosmology. Current instruments such as the Atacama Large Millimeter/submillimeter Array (ALMA) and the Hubble Space Telescope (HST) are being focussed on these epochs, providing detailed measurements of individual high-redshift objects. Ongoing molecular line surveys of small areas of sky (1--10 sq.\ arcmin) are constraining the properties of the brightest galaxies during the Epoch of Galaxy Assembly (for a  review, see \citealt{carilli_walter_13}), and planned instruments will extend the reach of such small-area surveys to the EoR, e.g.\ the next generation Very Large Array (ngVLA; \citealt{decarli_etal_18, walter_etal_19}), the James Webb Space Telescope (JWST; \citealt{finkelstein_etal_17}) and the Nancy Grace Roman Space Telescope (NGRST; \citealt{akeson_etal_19, koekemoer_etal_19}). On the other end of the scale, measurements of the cosmic microwave background (CMB) provide a constraint on the total reionization optical depth over the largest angular scales. 

What is lacking, however, is a way to bridge this huge range in spatial scales, providing constraints on the faintest galaxies that make up the bulk of the population while surveying large cosmic volumes in a reasonable time. Spectral line intensity mapping (LIM) is an emerging technique (see \citealt{kovetz_etal_17} for a review) that has the potential to fulfill this requirement and complement both galaxy surveys and constraints from the CMB. Unlike galaxy surveys, which trace the large-scale distribution of mass by individually detecting large numbers of galaxies, LIM measures the aggregate emission of spectral lines from unresolved galaxies and the inter-galactic medium (IGM) (Figure~\ref{fig:flowchart}). The redshift of the spectral line locates the emission in the line-of-sight direction. This allows efficient mapping of the cosmic luminosity density from a variety of spectral lines over a huge volume of the Universe.

Initially, most of the line intensity mapping effort focussed on 21-cm hydrogen emission from the neutral IGM, with several large experiments now underway (e.g., HERA, \citealt{deboer_etal_17}; CHIME, \citealt{bandura_etal_14}) or planned (e.g., HIRAX, \citealt{newburgh_etal_16}; SKA, \citealt{santos_etal_15}). Interest has rapidly grown in the use of this technique to trace galaxies using redshifted 21\,cm as well as the rotational carbon-monoxide (CO) transitions, the \Cii\ fine-structure line, and Ly$\alpha$, among others. 

Using CO as the tracer molecule for intensity mapping studies is complementary to other probes and has several advantages. Many observations, starting with \citet{brownvandenbout91}, have demonstrated that line emission from the CO transitions is bright even at high redshift. The CO emission from galaxies correlates strongly with the infrared luminosity (e.g.\ \citealt{carilli_walter_13}), a proxy for star formation, providing a picture that is complementary to studies of the neutral IGM. The multiple emission lines of CO, with a well defined frequency spacing, enable the signal to be separated from contaminating signals; this feature is not available to either  \Hi\  or \Cii\ measurements. The levels of foreground contamination in a CO survey are substantially lower than for many other types of line intensity mapping, as CO suffers from neither the exceptionally high levels of continuum foregrounds present in 21-cm surveys nor the bright spectral line foreground present in \Cii\ surveys.

Purpose-built pathfinder-scale experiments are currently pursuing detections from the ground in a variety of emission lines, including CO (COMAP\footnote{\url{http://comap.caltech.edu}}, the subject of this paper; AIM-CO, \citealt{kovetz_etal_17}) and \Cii\ (EXCLAIM, \citealt{Cataldo_etal_21}; TIM, \citealt{aguirre_18}; TIME, \citealt{crites_etal_14}; FYST, \citealt{CCAT_21}; and CONCERTO, \citealt{lagache_18}). Observations using facility instruments (e.g.\ ALMA, CARMA, VLA) have yielded intensity mapping constraints on the CO shot-noise power spectrum and on the CO--galaxy cross-spectrum at redshifts in the range $1 < z < 5$ \citep{keating_etal_16, keating_etal_20, Keenan21}. SPHEREx \citep{korngut_etal_18}, a NASA MIDEX-class mission to perform spectroscopic surveys in multiple lines, including H$\alpha$, H$\beta$ and Ly$\alpha$, has been selected by NASA for launch in 2023. 

COMAP was funded by the United States' National Science Foundation (NSF) in 2015 to construct a pathfinding instrument for CO LIM. This Pathfinder is the first step in a long-term program that ultimately aims to trace the evolution of the global properties of galaxies, through the epoch of their assembly to cosmic reionization. Given the nascent state of the line intensity mapping field and the challenges involved in targeting the uncertain CO signal from the EoR, the goals for the Pathfinder focused on i) constraining the CO clustering power spectrum of galaxies at $z \sim 3$ and ii) demonstrating the technology and techniques needed for observations targeting the EoR. An ancillary science goal was to perform continuum observations of Galactic and extragalactic targets.

\begin{table*}
\centering
\caption{COMAP Early Science Papers.}
\label{tab:papers}
\begin{tabular}{llll}
\hline
\hline
 & Title & Reference & In this work \\
\hline
I. & Overview & (this paper) &  \\
II. & Pathfinder Instrument & \citet{es_II} & \S~\ref{sec:instrument} \\
III. & CO Data Processing & \citet{es_III} & \S~\ref{sec:analysis}\\
IV. & Power spectrum methodology and results & \citet{es_IV} & \S~\ref{sec:spectra}\\
V. & Constraints and Forecasts at $z\sim3$ & \citet{es_V} & \S~\ref{sec:modeling}, \S~\ref{sec:constraints}, \S~\ref{sec:zeq3_forecasts}\\
VI. & The COMAP Galactic Plane Survey & \citet{es_VI} & \S~\ref{sec:gps}\\
VII. & Prospects for CO Intensity Mapping at Reionization & \citet{es_VII} & \S~\ref{sec:zeq6_forecasts} \\
\hline
\end{tabular}
\end{table*}

\begin{figure}[t!]
\centering
\includegraphics[width=0.45\textwidth]{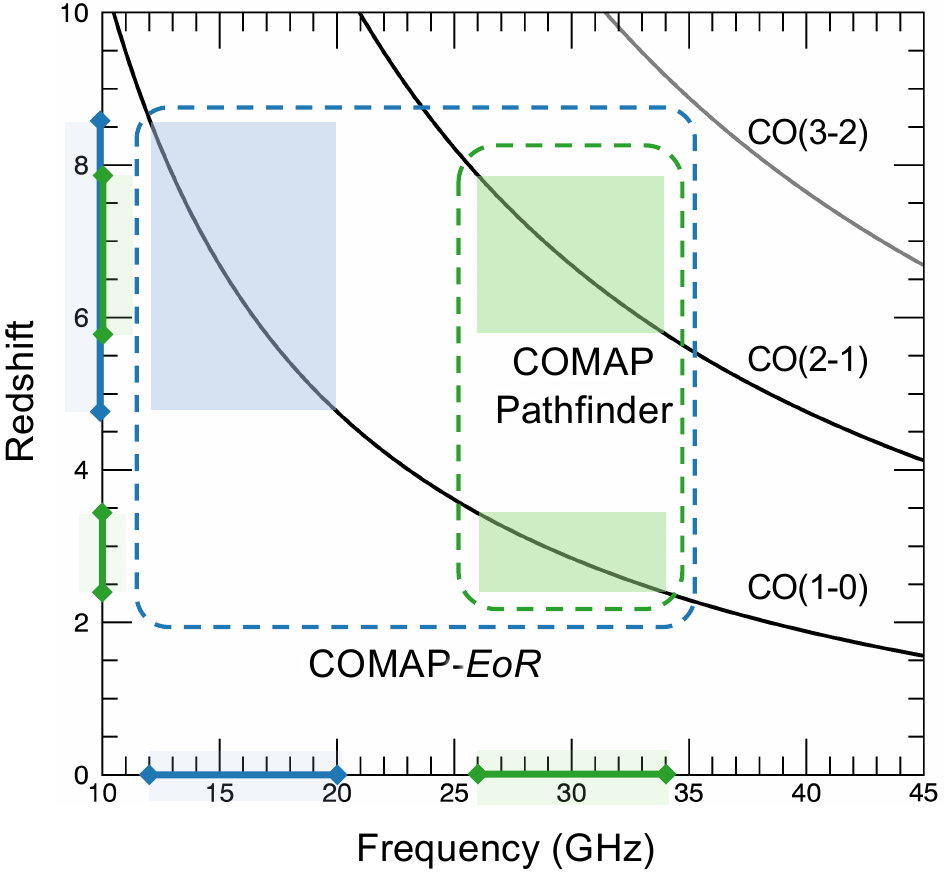}
\caption{Redshift of the three lowest CO transition lines as a function of observed frequency. The frequency coverage for the COMAP Pathfinder Survey (26--34\,GHz) is sensitive to the \mbox{CO(1--0)} line in the redshift range $z=2.4$--$3.4$ and the CO(2--1) line at \mbox{$z=6$--$8$}. Also shown is the frequency coverage of a future \comapeor\ survey, in which a second frequency channel (12--20\,GHz) is added, sensitive to the CO(1--0) line at $z=4.8$--$8.6$.
}\label{fig:colines}
\end{figure}

In this paper, the first of a series, we present an overview of early results, starting with a description of the Pathfinder instrument in \S\ref{sec:instrument}. For the LIM science (\S\ref{sec:lim_survey}), we describe the observations conducted during the Pathfinder's first LIM observing season (\S\ref{sec:observations}) and the LIM analysis pipeline (\S\ref{sec:analysis}). Next, we present the power spectrum results from this first season (\S\ref{sec:spectra}) and discuss their implications for the global properties of galaxies at $z \sim 3$ (\S\ref{sec:modeling}). For the continuum science, we present early results from a 30\,GHz survey of the Galactic plane (\S\ref{sec:gps}). We conclude by presenting forecasts for the nominal 5-year Pathfinder survey (\S\ref{sec:zeq3_forecasts}), and prospects for the next phases of the project, targeting galaxies at $z=5$--$9$ (\S\ref{sec:zeq6_forecasts}). The other papers in the series (see Table~\ref{tab:papers}) describe these and other aspects in significantly more detail.

\begin{figure*}[t!]
\centering
\includegraphics[width=0.99\linewidth]{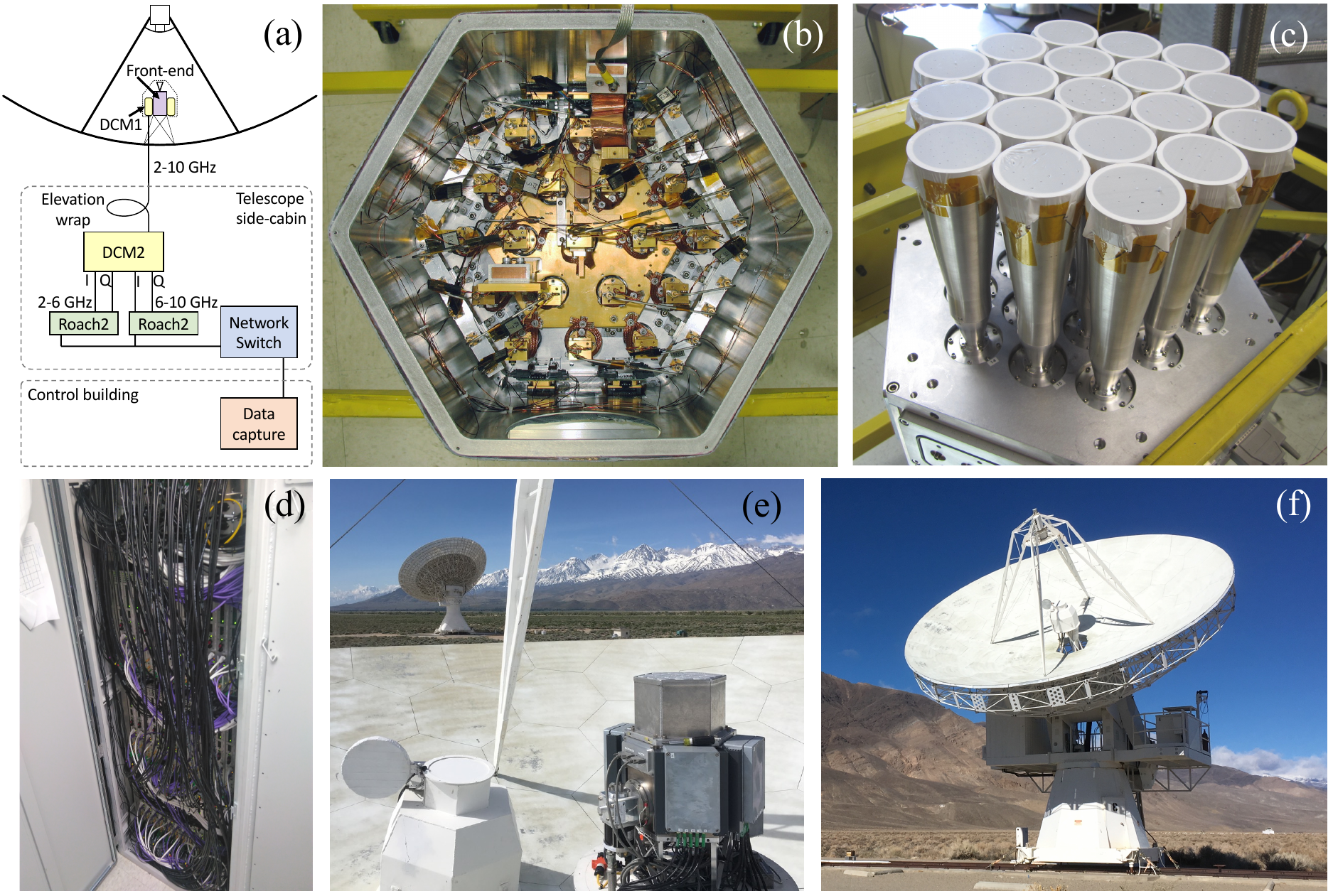}
 \caption {\label{fig:system} {\it\bf (a)} A block diagram showing the signal path for a single feed of the 19-feed Pathfinder receiver, consisting of the cryogenic front end, first- (DCM1) and second- (DCM2) stage downconverters, ROACH-2 spectrometers and the data storage computer.  {\it\bf (b)} Cryostat interior showing 19 low-noise amplifiers.  {\it\bf (c)} Cryostat exterior showing 19 feed horns.  {\it\bf (d)} Digital rack in telescope side-cabin with 38 ROACH-2 spectrometers.  {\it\bf (e)} Cryostat and DCM1 modules mounted at the secondary focus, with weather enclosure removed.  {\it\bf (f)}  The Pathfinder receiver fielded on a 10.4-m telescope at OVRO.}\label{fig:hardware}
\end{figure*}

\section{The Pathfinder Instrument} \label{sec:instrument}

The Pathfinder targets the 26--34\,GHz frequency range, which is sensitive to the 115.27\,GHz CO(1--0) line in the redshift range $z=2.4$--$3.4$ and the 230.54\,GHz CO(2--1) line at $z=6$--$8$ (see Figure~\ref{fig:colines}). The Pathfinder receiver (Figure~\ref{fig:hardware})  is a single-polarization 19-feed focal plane array, deployed on a 10.4-m Cassegrain telescope \citep{leighton_final_1977} at the Owens Valley Radio Observatory (OVRO), resulting in a resolution of 4.5\,arcmin at 30\,GHz. Table~\ref{tab:instrument} summarizes the main instrument parameters. The first down-conversion stage, located in modules mounted on the exterior of the cryostat, shifts the 26--34\,GHz RF band to a first intermediate frequency ($1^{\text{st}}$ IF) of 2--10\,GHz. The second down-conversion stage occurs inside the telescope side-cabin, where the 2--10\,GHz band from each feed is split into two 4\,GHz-wide bands, each of which is quadrature down-converted to produce an ``in-phase'' ($I$) and ``quadrature'' ($Q$) signal at a $2^{\text{nd}}$ IF of 0--2\,GHz. Each $IQ$ pair is input to a ``ROACH-2'' FPGA-based spectrometer (from a hardware design by the Center for Astronomy Signal Processing and Electronics Research (CASPER); \citealt{parsons_etal_06}). Custom FPGA code in each ROACH-2 performs separation into upper (USB) and lower sidebands (LSB), resulting in four 2\,GHz-wide sidebands from each feed, each of which has 1024 spectral channels (i.e.\ $\sim2$\,MHz spectral resolution). In order to process the 8\,GHz bandwidth from each of the 19 feeds, 38 ROACH-2 spectrometers are needed. Spectra are recorded every 20\,ms and sent via Ethernet to a storage machine in a nearby control building. From there, these data are transmitted to the Caltech campus via internet and stored on disk arrays. The spectra are then combined with pointing and housekeeping data and are available for further analysis and processing. 

\begin{table}
\centering
\caption{Pathfinder instrument parameters.}
\label{tab:instrument}
\begin{tabular}{lc}
\hline
\hline
Parameter & Value \\
\hline
Primary reflector diameter & 10.4 m \\
No.\ of feeds & 19 \\
Polarization & left-circular \\
Beam FWHM @ 26, 30, 34\,GHz & 4.9, 4.5, 4.4 arcmin \\
Beam separation on sky  & 12.0 arcmin (center to center) \\
System temperature\tablenotemark{\footnotesize{a}}  & 34--60\,K (median 44\,K) \\
Frequency resolution (native) & 1.953125\,MHz \\
Frequency resolution (science) & 31.25\,MHz\\
\uline{Frequency bands} & \uline{$1^{\text{st}}$ IF (RF) frequency}\\
A LSB & 2--4\,GHz (26--28\,GHz) \\
A USB & 4--6\,GHz (28--30\,GHz) \\
B LSB & 6--8\,GHz (30--32\,GHz) \\
B USB & 8--10\,GHz (32--34\,GHz) \\
\hline
\end{tabular}
\tablenotetext{a}{\footnotesize{The range given corresponds to the central 95\% of all scans between 35--65 degrees elevation, for all feeds and all sidebands (see \citealt{es_IV}, Appendix~A).}}
\end{table}

A ``calibration vane,'' consisting of microwave absorber material at ambient temperature, is mounted on the side of the receiver (see Figure~\ref{fig:hardware}, (e) and (f)) . This vane can be  moved under computer control into the path of the feed horns in order to present a blackbody at ambient temperature for determination of the system temperature, which is used to calibrate the data to brightness temperature units.

When not in use, the vane rests at a position in the plane of the secondary supports so that it causes no additional shadowing of the aperture.

\section{The Pathfinder Line Intensity Mapping Survey} \label{sec:lim_survey}

\subsection{Field Selection} \label{sec:selection}

The primary goal of the Pathfinder is to detect the power spectrum of CO(1--0) fluctuations from galaxies at $z=2.4$--$3.4$ on scales relevant to clustering; that is, $\gtrsim 10$\,Mpc, corresponding to spatial Fourier modes, $k\lesssim0.6\,{\rm Mpc}^{-1}$ (see Figure~\ref{fig:flowchart}, right). The total detection significance (over all observable $k$) is predicted (in a model-dependent way) to be optimal for an individual field size $\leq$ 1 deg$^{2}$ \citep{Breysse_etal_14}. However, the necessity to scan the telescope \citep{es_III} and the size of the feed array on the sky both impose a practical limit on how small each field can be, resulting in an effective size of around 4 deg$^2$. Field selection was influenced by the following considerations: i) fields were distributed in right ascension in order to maximize observing efficiency, ii)  field locations were selected to overlap with the Hobby-Eberly Telescope Dark Energy Experiment (HETDEX) galaxy survey \citep{hill_etal_21,gebhardt_etal_21, hill_etal_08} in order to cross-correlate with their Ly-$\alpha$ emitter catalog, and iii) bright 30\,GHz point sources were avoided.

\begin{table}
\centering
\caption{COMAP Season 1 field location, effective area and effective integration time. The effective area corresponds to a cutoff in the associated map's white noise level equivalent to 95\% of the total integration time. The effective integration time is that assuming 100\% acceptance of all 19 feeds and frequency channels.}
\label{tab:patches}
\begin{tabular}{lcccc}
\hline
\hline
& & &$\Omega_{\rm eff}$  &$\tau_{\rm eff}$ \\
Name & RA (J2000) & Dec (J2000) & (deg$^2$) & (h) \\
\hline
Field 1& 01$^{\rm h}$ 41$^{\rm m}$ 44$\fs$4&00$\degr$ 00$\arcmin$ 00$\farcs$0& 4.24& 303.7\\
Field 2& 11$^{\rm h}$ 20$^{\rm m}$ 00$\fs$0&52$\degr$ 30$\arcmin$ 00$\farcs$0& 3.54& 346.5\\
Field 3& 15$^{\rm h}$ 04$^{\rm m}$ 00$\fs$0&55$\degr$ 00$\arcmin$ 00$\farcs$0& 4.00& 245.9\\
\hline
\end{tabular}
\end{table}


\begin{figure*}
\centering
\includegraphics[width=\linewidth]{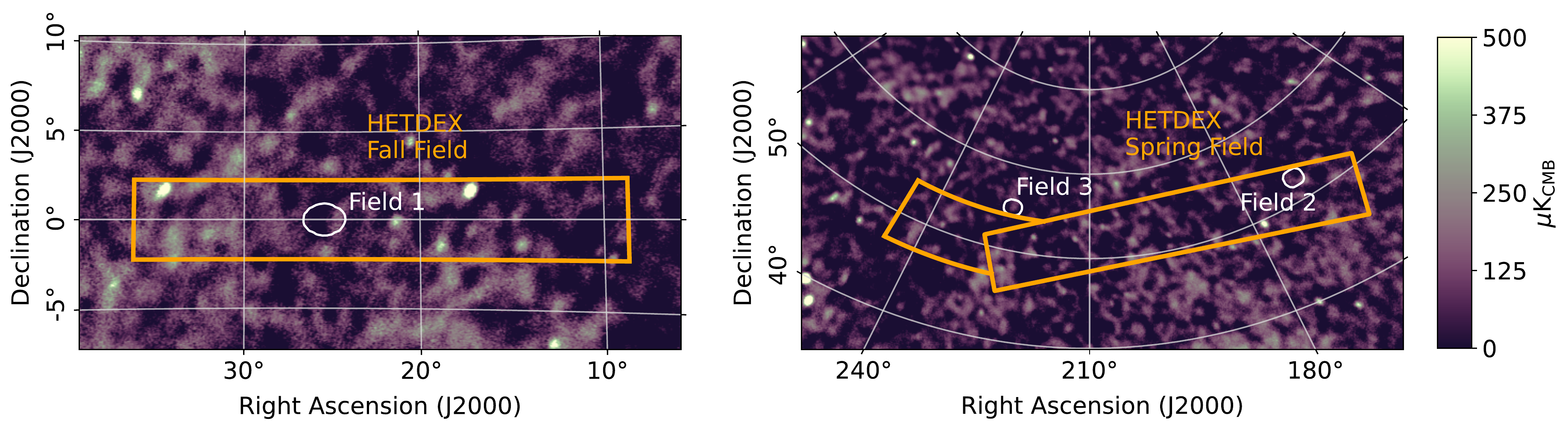}
\caption{The three CO fields, overlaid on the Planck LFI 30\,GHz full-mission map \citep{Planck2020}. The white contours indicate the $\sim4$\,deg$^2$ coverage of each field. The orange outlines correspond to the HETDEX Fall (left) and Spring (right) fields \citep{gebhardt_etal_21}. \label{fig:patches}}
\end{figure*}

The result of these considerations was a set of three \mbox{$\sim$\,4-deg$^2$} fields as shown in Table~\ref{tab:patches} and Figure~\ref{fig:patches}.

\subsection{Observations} \label{sec:observations}

Observations using the Pathfinder began in 2019 June and continued until 2020 August, when the receiver was removed from the telescope for  maintenance. Observing resumed in 2020 November and we refer to the period before receiver maintenance as ``Season 1''. The LIM results presented here and in the other Early Science papers of this sequence are all based on data from Season~1 only. LIM data taken after this will be included in future publications. 

Each observation of a LIM field consists of several scans, where one scan is the period between two re-pointings of the telescope, during which the telescope performs the same motions around a fixed point in azimuth and elevation while the target field drifts through. Two main scan patterns are used: i) slewing in azimuth, at a fixed elevation (constant elevation scan, or ``CES'') and ii) a Lissajous pattern.  Observations alternated between CES and Lissajous scan patterns on a daily basis during Season 1.

During an observation, spectra are recorded every 20\,ms and written to files after a period typically 1~h in length. A measurement of the system temperature ($T_{\rm sys}$) using the calibration vane is performed at the beginning and end of this period.

Table~\ref{tab:patches} shows the effective integration time obtained on each field after pipeline filtering, for an ideal instrument with 100\% acceptance of all feeds and frequencies channels.

\begin{figure}[t!]
\centering
\includegraphics[width=0.99\linewidth]{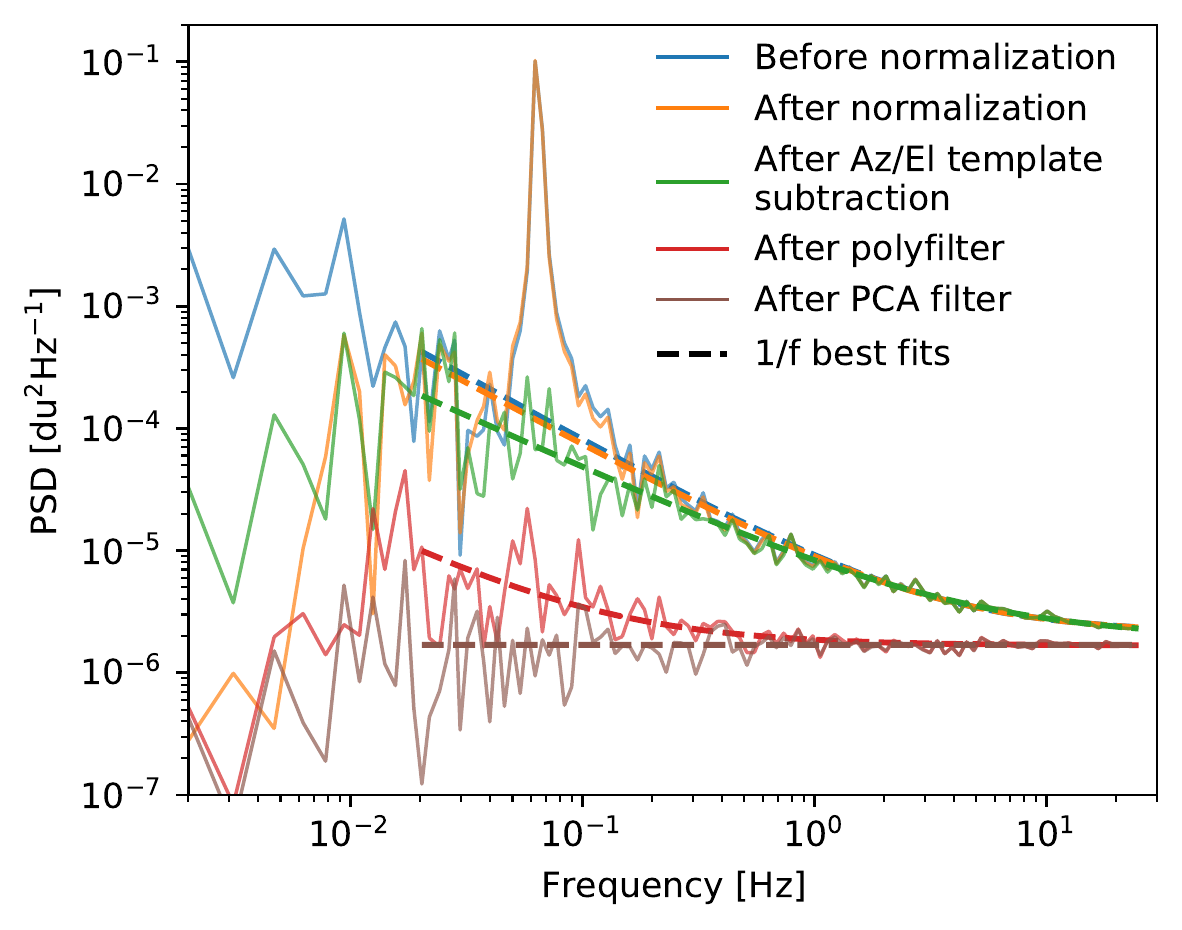}
 \caption{Power spectrum of a single scan from a 31.25\,MHz band around 26.2\,GHz at different stages in the analysis pipeline, with $1/f$ noise curves fitted. The power spectrum is binned with logarithmic bin-sizes towards higher frequencies for clarity. Lower frequencies have been excluded from the fit, as these scales are greatly suppressed at the normalization stage. The end result of the various filters is a flat power spectrum indicating a very low level of correlated noise. See \citet{es_III} for details of the filters involved. \label{fig:1f_fits}}
\end{figure}

\subsection{Data Analysis} \label{sec:analysis}

Data processing starts with the raw data as recorded by the instrument, together with pointing information and house-keeping data. The first processing step is to partition each observation into individual scans, based on pointing information. The next step is to process the time-ordered data (TOD) by applying a series of filters and a time-varying gain normalization, as follows (see Figure~\ref{fig:1f_fits} for the effect of each step).

\begin{itemize}
\item The raw TOD for each frequency channel are divided by their running mean and shifted to zero-mean by subtracting 1.0 from the result. This normalization removes slow drifts and effectively flattens the instrument passband. 
\item Next, contaminating signals that are correlated with pointing in the telescope frame (e.g.\ ground pickup) are removed by fitting for and removing contributions that vary with azimuth and $1/\sin(\rm{elevation})$. This works well for CES scans but we have found that Lissajous-scanned data are more prone to residual ground pickup. As a result, we only include CES scans in our Season 1 analysis and observations after Season 1 only use CES scans.
\item To the resulting TOD, we then perform a linear fit to all frequency channels within a single sideband, at each time step. This ``polyfilter'' removes all signals that are common across frequencies, including receiver gain fluctuations, but also contaminating signals that vary smoothly in frequency, such as the atmosphere, cosmic microwave background and Galactic foregrounds. This step reduces the noise level by over an order of magnitude and the filtered data are correlated at only a few per cent. 
\vspace{-3pt}
\item A principal component analysis (PCA) identifies signals common across the focal plane and across frequencies then removes the leading modes.
\item Individual frequency channels are flagged based on diagnostics such as channel-to-channel correlations. We also remove channels where out-of-band signals are aliased. 
\end{itemize}

Calibration in brightness temperature units is then achieved by multiplying the cleaned and co-added TOD by the system temperature measured using the calibration vane.

Data selection is performed on the basis of a variety of statistics that track issues ranging from poor weather conditions to elevations affected by ground pickup. After these cuts have been imposed, we average groups of 16 frequency channels and produce maps for each of these coarser 31.25\,MHz-wide channels in order to match the expected width of the observed CO lines \citep{chung_etal_19}.

Figure~\ref{fig:1f_fits} shows the cumulative effect of each of the filters described above; the end result is a dataset with a very low level of correlated noise without removing significant levels of signal on the scales of interest (see discussion in next section).

\subsection{CO Power Spectra} \label{sec:spectra}

The analysis pipeline produces a three-dimensional (3D) map of CO temperature for each field, from which we wish to estimate the ``auto-power spectrum'', or the variance of the map's Fourier components as a function of spatial frequency. Estimation of this auto-power spectrum for an experiment such as COMAP is challenging. The 3D maps are not uniform in sensitivity and so estimates of the noise power spectrum, which must be subtracted off the measured power spectrum, will be biased by the parts of the maps with the highest noise level. A ``pseudo-spectrum'' analysis \citep{hivon2002} allows us to weight the lower-noise portions of the maps most highly. Even so, the expected level of the CO signal is just a few microkelvin, compared to a typical system temperature of 44\,K and contaminating signals of a few millikelvin. To detect such a weak signal reliably, we need a power-spectrum estimation method that is robust in the presence of systematic errors.

The design of the COMAP Pathfinder instrument has some features that naturally lend themselves to such a method. Each of the 19 receiver chains is independent and so a ``cross-spectrum'' formed from maps computed for different feeds (i.e.\ the covariance between the Fourier components of the two maps) will be insensitive to systematic errors that are peculiar to any of the individual feeds. The cross-spectrum has the further advantage that the noise contributions of the two different maps do not contribute to its mean and so it directly provides an unbiased estimate of the signal spectrum.

While some systematic errors will be common to all feeds, the cross-spectrum will still be insensitive to such errors as long as the cross-correlation is across not only different feeds, but also different subsets of the data that are selected by variables related to these errors. An example of one important systematic error relevant to COMAP is ground pickup. In this case, we can divide the data in two according to scan elevation. We can then ensure that we never form a cross-spectrum using data from the same elevation subset, so that the resulting spectra are as insensitive as possible to this form of contamination. We refer to the power spectrum estimated in this way as the feed--feed pseudo-cross spectrum (FPXS).

By using only cross-power products from two independent halves of the data, we lose at least a factor of $\sqrt{2}$ in sensitivity compared to the auto-power spectrum. We can however approach the auto-power spectrum sensitivity by splitting the data into many more independent parts and taking the cross-spectra of all possible combinations. 

The final pseudo-power cross-spectrum estimate is formed by taking the noise-weighted average of all the individual cross-power spectra. Uncertainties in this power spectrum are estimated by assuming i) that they are dominated by the error bar on the noise power spectrum and ii) that the noise is only uncorrelated white noise. We can then create a large number of simulated noise maps, where the value in each voxel is drawn from a zero-mean distribution with a standard deviation given by the noise map in that voxel. The standard deviation of the resulting power spectra in each $k$-bin then gives an estimate of the uncertainty in the signal power spectrum in that bin.

\begin{figure}
	\begin{center}
		\includegraphics[width=0.49\textwidth]{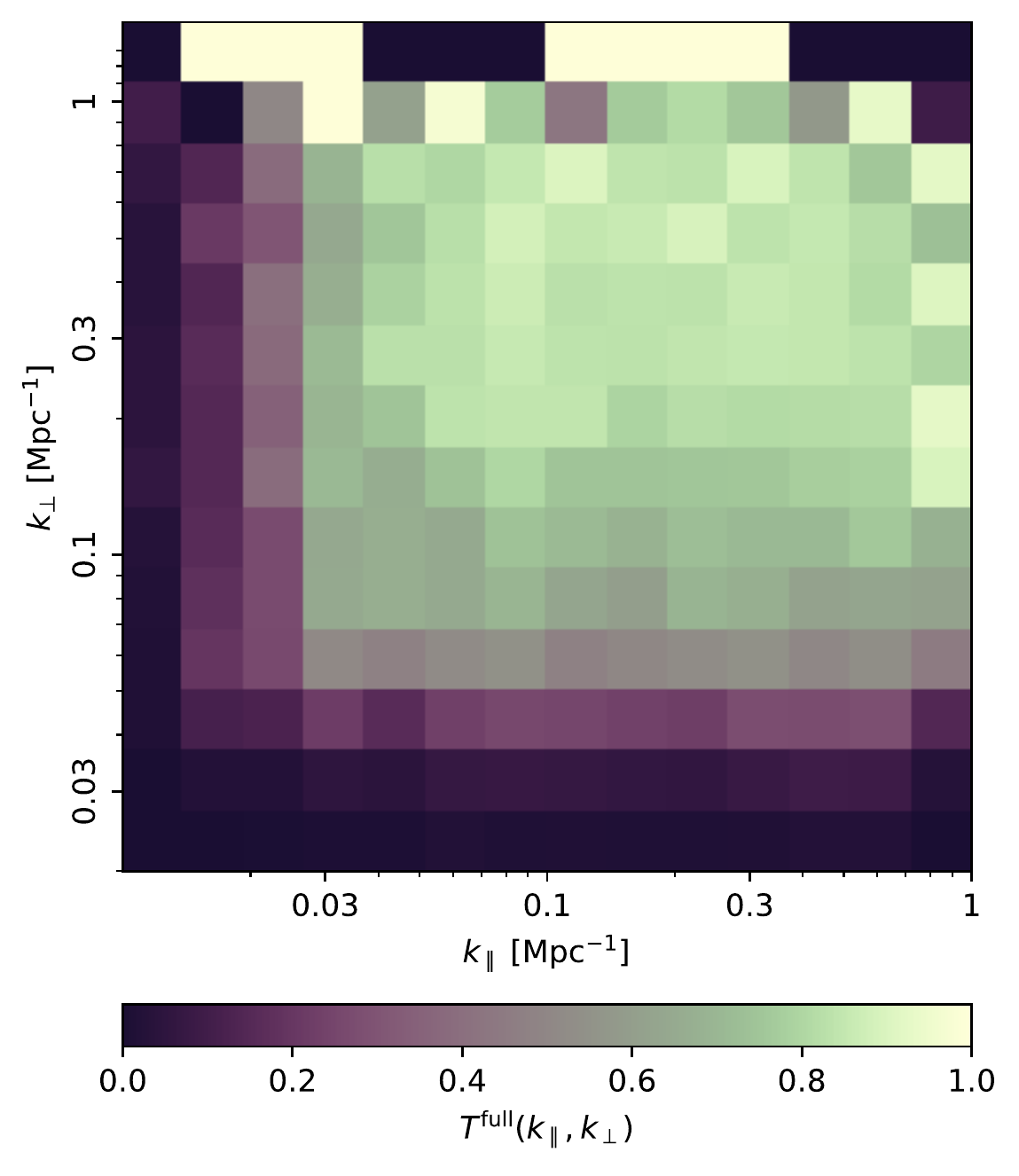}
		\caption{Pipeline transfer function for the cylindrically averaged power spectrum, based on a single signal realization and roughly three hours of data. \label{fig:TF_2d_CES}}
	\end{center}
\end{figure}

The effects on the signal of the instrumental beam and of pipeline filtering, calibration and map-making are taken into account using transfer functions. The beam transfer function is estimated by comparing the power spectra of mock signal-only maps to those for the maps smoothed with a beam model. We derive the beam model from a physical optics simulation \citep{es_II}, normalized using astrophysical sources to account for the main beam efficiency. 

The pipeline transfer function is estimated by combining (noise-dominated) raw data with a simulated signal-only time ordered data-set and processing it through the analysis pipeline in an identical manner to the raw data. Taking the power spectrum of this processed combination and subtracting off that of the raw data, we estimate the transfer function by comparing the result with the power spectrum of the signal-only simulation. We can see that the effect of the pipeline filters (Figure~\ref{fig:TF_2d_CES}) is to suppress sensitivity mostly on the largest scales, both along and perpendicular to the line of sight, while retaining sensitivity on the clustering scales that are of most interest. 

\begin{figure}
	\begin{center}
		\includegraphics[width=0.47\textwidth]{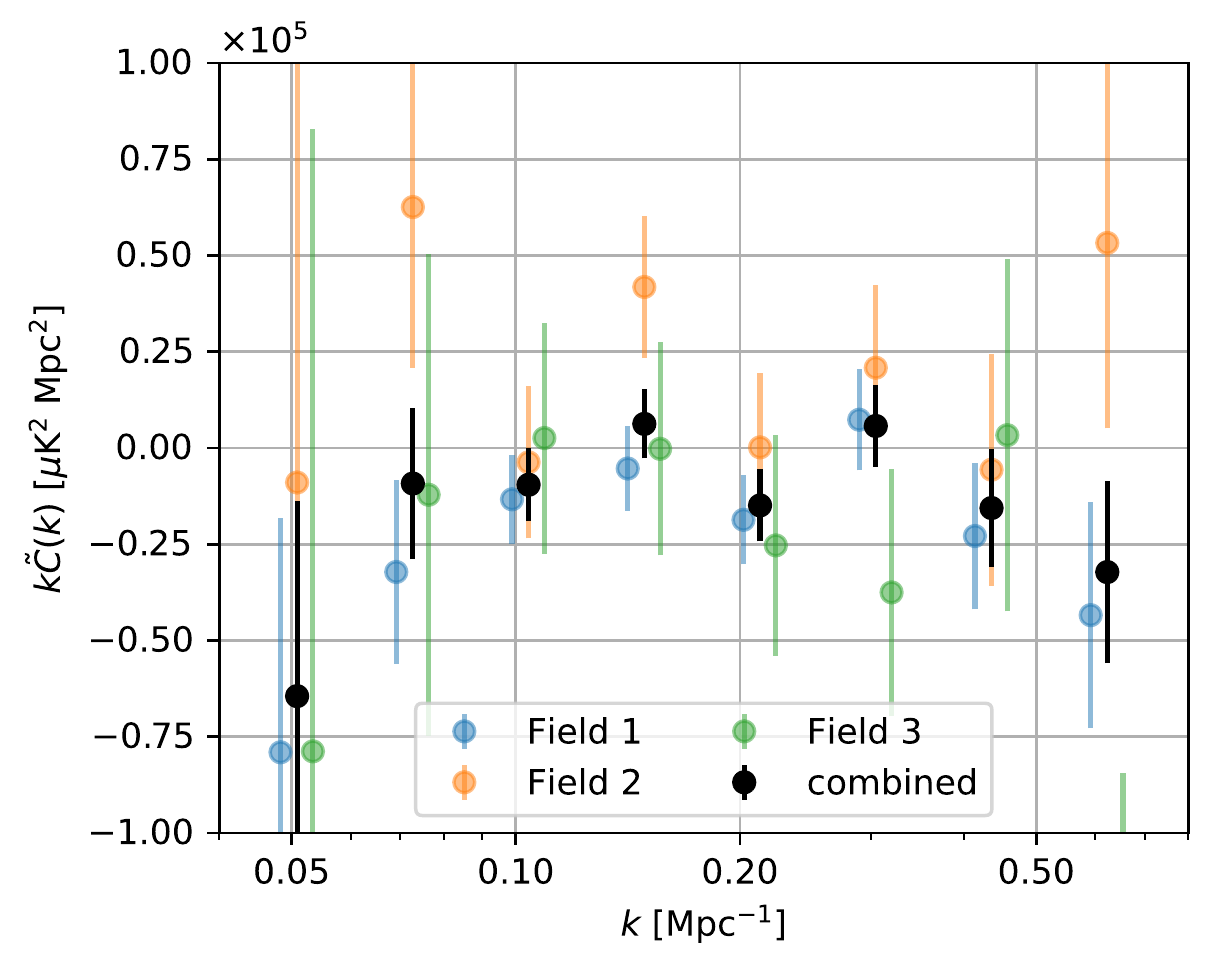}
		\caption{Spherically averaged mean pseudo-cross spectra for Season 1 observations of Field 1 (blue), Field 2 (orange) and Field 3 (green). These spectra were generated from all the accepted data using the FPXS statistic. In addition the full transfer function has been applied, to de-bias the signal estimate. Data points from the different fields are offset slightly in $k$ from their actual values to make them easier to distinguish. \label{fig:FPXS_1d}}
	\end{center}
\end{figure}

With these transfer functions in hand, we produce an unbiased estimate of the signal power spectrum by dividing the pseudo-power cross-spectrum estimate by the full transfer function (combining beam and pipeline effects). Figure~\ref{fig:FPXS_1d} shows the final Season 1 power spectra for each field. As a consistency check, we also form cross-spectra between maps of our different fields and find them consistent with  expectations for white noise (see \citealt{es_IV} for details). 

\subsection{Expectations for the CO signal at \texorpdfstring{$z\sim3$}{z ~ 3}}\label{sec:modeling}

A model for the expected CO signal at $z\sim3$ was presented by \cite{li_etal_16} and used to predict the significance of a detection for a COMAP-like experiment. Since then, a number of new observational constraints have become available and we have fielded the Pathfinder, thereby gaining an understanding of the instrument's real-world sensitivity. These developments have provided a motive to update the fiducial model (and also the Pathfinder forecasts, which we discuss in \S~\ref{sec:zeq3_forecasts}).

\begin{figure*}
    \centering
    \includegraphics[width=1.0\linewidth]{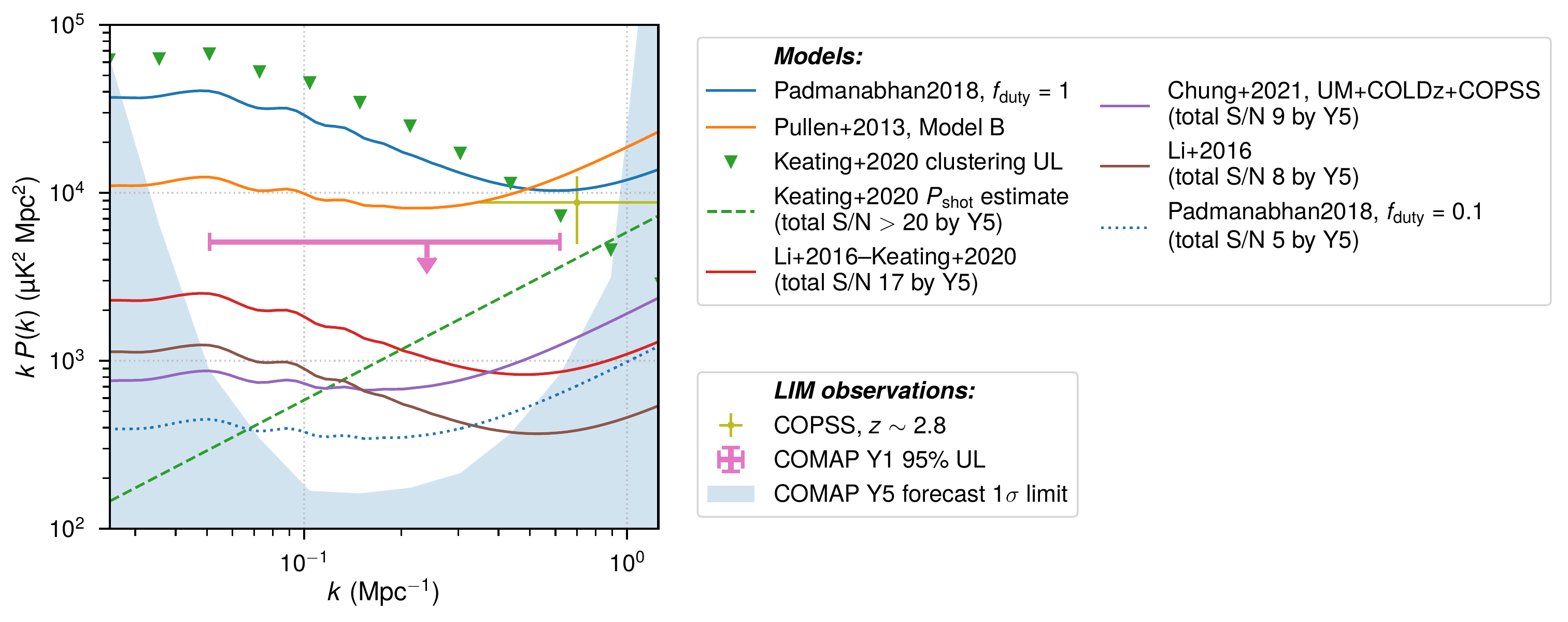}
    \caption{COMAP Pathfinder Season 1 constraint (pink) on the redshift-space CO(1--0) power spectrum at $z\sim3$, alongside the predictions from various models and our Year 5 Pathfinder sensitivity forecast (blue shaded area). The models include i) the fiducial COMAP data-driven model from \cite{es_V} (``UM+COLDz+COPSS''), ii) an alternative model from \cite{keating_etal_20} (``Li+2016--Keating+2020'') with emission from faint galaxies that may be missed by the surveys informing the fiducial model, and iii) models based on $L(M_h)$ relations from~\cite{padmanabhan_18},~\cite{Pullen13}, and~\cite{li_etal_16}. We also show the COPSS result \citep{keating_etal_16} as a direct $P(k)$ measurement and as a constraint on the clustering and shot-noise amplitudes \citep{keating_etal_20}. For each, the legend indicates the expected signal-to-noise ratio with which we would reject the null hypothesis (i.e., excluding sample variance from the calculation).}
    \label{fig:Pk_mods}
\end{figure*}

The updated model follows the general prescription of \cite{li_etal_16}, using a ladder of relations to assign CO luminosity to halo mass, as follows. A power law is assumed to relate CO ($L'_{\rm CO})$ to infrared ($L_{\rm IR}$) luminosity, which is considered in turn to be proportional to star-formation rate (SFR). We connect average SFR to halo mass using empirically-constrained ``UniverseMachine'' (UM) modeling by \cite{Behroozi19}. The combination of these relations has the form of a double power law (motivated from data-driven treatments in \citealt{padmanabhan_18} and \citealt{moster2010}) and is a function of their seven individual parameters, some of which are degenerate. In a change to the approach adopted by \cite{li_etal_16}, we therefore simplify the combined relation between CO luminosity, $L'_{\rm CO}$, and halo mass, $M_h$, to one with four effective free parameters as follows:
\begin{equation}\frac{L'_{\rm CO}(M_h)}{{\rm K\,km\,s}^{-1}\,{\rm pc}^2} = \frac{C}{(M_h/M)^A+(M_h/M)^B}\equiv\frac{C}{m^A+m^B}.\end{equation} 
(see \citealt{es_V} for expressions relating $A, B, C$ and $M$ to those of the original scaling relations). To these four parameters, we add $\sigma$, which accounts for scatter in the halo/SFR and $L_{\rm IR}$/$L'_{\rm CO}$ relations.

Priors on the values for these parameters were propagated from those on $L_{\rm IR}$-$L'_{\rm CO}$ and $L_{\rm IR}$-SFR used by \cite{li_etal_16} and the 68\% interval around the best-fit values of the other parameters from \cite{Behroozi19}. We also incorporated constraints based on recent observations of CO(1--0) emission from galaxies around $z\sim3$. These include i) the CO Luminosity Density at High-$z$ (COLD$z$) survey \citep{pavesi_etal_18, riechers_etal_18}, a blind molecular line survey of the COSMOS and GOODS-N fields, and ii) the CO Power Spectrum Survey \citep[COPSS;][a re-analysis of archival Sunyaev-Zel'dovich Array (SZA) data]{keating_etal_16}. Information from these observations is incorporated into our priors by running a Markov Chain Monte Carlo (MCMC) sampler with initial priors on the five parameters, converting halo masses from the Bolshoi-Planck simulation \citep{behroozi_etal_19} into CO luminosities and comparing the calculated luminosity functions and shot noise with those based on the observational data. We use the posterior distribution of the model parameters to generate an updated fiducial estimate (``UM+COLDz+COPSS'') of the CO power spectrum.

Figure~\ref{fig:Pk_mods} shows our fiducial model along with several others from the literature. It can be seen that our model is somewhat low compared to the constraint from COPSS. This highlights a potential limitation of our approach. Its reliance on constraints from a galaxy survey, which can only be sensitive to the brightest objects, means it may omit a significant fraction of the very signal we are searching for. If there is a significant contribution from faint galaxies to the overall average CO luminosity, as hinted by the available LIM constraints (such as those from COPSS), then the fiducial model can only be interpreted as a lower limit. 

For this reason, we also consider a model that does not suppress the contribution from faint sources \citep{keating_etal_20}. Like the fiducial model, this takes the general approach of~\cite{li_etal_16}, but uses newer (albeit exclusively local) IR--CO correlation fits from~\cite{Kamenetzky16}. The power spectrum for this model is shown in Figure~\ref{fig:Pk_mods} labeled ``Li+2016--Keating+2020''. For our sensitivity forecasts, discussed in \S~\ref{sec:future}, we adopt both the UM+COLDz+COPSS and Li+2016--Keating+2020 models in order to represent the range of possible CO signals.

\subsection{Constraints from Season 1 Pathfinder Observations\label{sec:constraints}}

Pipeline, observing and operational improvements will increase our sensitivity significantly in subsequent observations. However, we already find that Season 1 data alone exclude two models for the power spectrum from the literature and provide the best LIM clustering constraint on CO(1--0) at $z\sim3$.

From our Season~1 data, we obtain a 95\% upper limit of $P_\mathrm{CO}(k) = -2.7 \pm 1.7 \times 10^4\mu\textrm{K}^2\,\mathrm{Mpc}^3$ on scales $k=0.051$--$0.62 \,\mathrm{Mpc}^{-1}$, or $kP_\mathrm{CO}(k)<5.1\times10^3$\,$\mu$K$^2$\,Mpc$^2$ at $k=0.24$\,Mpc$^{-1}$. This is the first direct 3D constraint on the clustering component of the CO(1--0) power spectrum. We plot this constraint in~\autoref{fig:Pk_mods} along with that from the COPSS re-analysis of~\cite{keating_etal_20} as well as several models. Our Season 1 upper limit excludes one of the model predictions of~\cite{padmanabhan_18} and Model B of~\cite{Pullen13} at 95\% confidence. Being entirely empirical, the former model is driven to a very high clustering amplitude by the incorporation of the COPSS constraint. Its exclusion demonstrates the importance of being able to directly constrain the clustering power spectrum. Exclusion of the latter model has implications for the applicability of the CO--SFR relation used to derive its predictions for CO luminosity at $z\sim3$.

Our upper limit is not yet sensitive enough to constrain the parameters of our fiducial model but we can use it to set limits on the possible contributions of clustering and shot noise to the total power spectrum, as follows. The observed CO power spectrum, $P_\mathrm{CO}(k)$, is the sum of clustering and shot-noise terms, 
\begin{equation}
    P_\mathrm{CO}(k) = A_{\text{clust}}P_m(k) + P_{\text{shot}},
\end{equation}
where $A_{\text{clust}}$ is the clustering amplitude, $P_m(k)$ is the dark matter power spectrum, and $P_{\text{shot}}$ is the scale-independent shot noise. (These components are shown schematically in Figure~\ref{fig:flowchart}, right). Mapping in redshift space imposes distortions on the observed intensity field, such that the clustering amplitude is given by $A_{\text{clust}}\approx \avg{T}^{2}(b^2+2b/3+1/5)$ for small $k$, where $\avg{T}$ is the mean CO line intensity and $b$ is the luminosity-weighted bias. Our data bounds $A_{\text{clust}}<66\,\mu$K$^2$, an order-of-magnitude improvement on the~\cite{keating_etal_20} COPSS re-analysis upper limit of 420\,$\mu$K$^2$.

\begin{figure}[t!]
    \centering
    \includegraphics[width=0.96\linewidth]{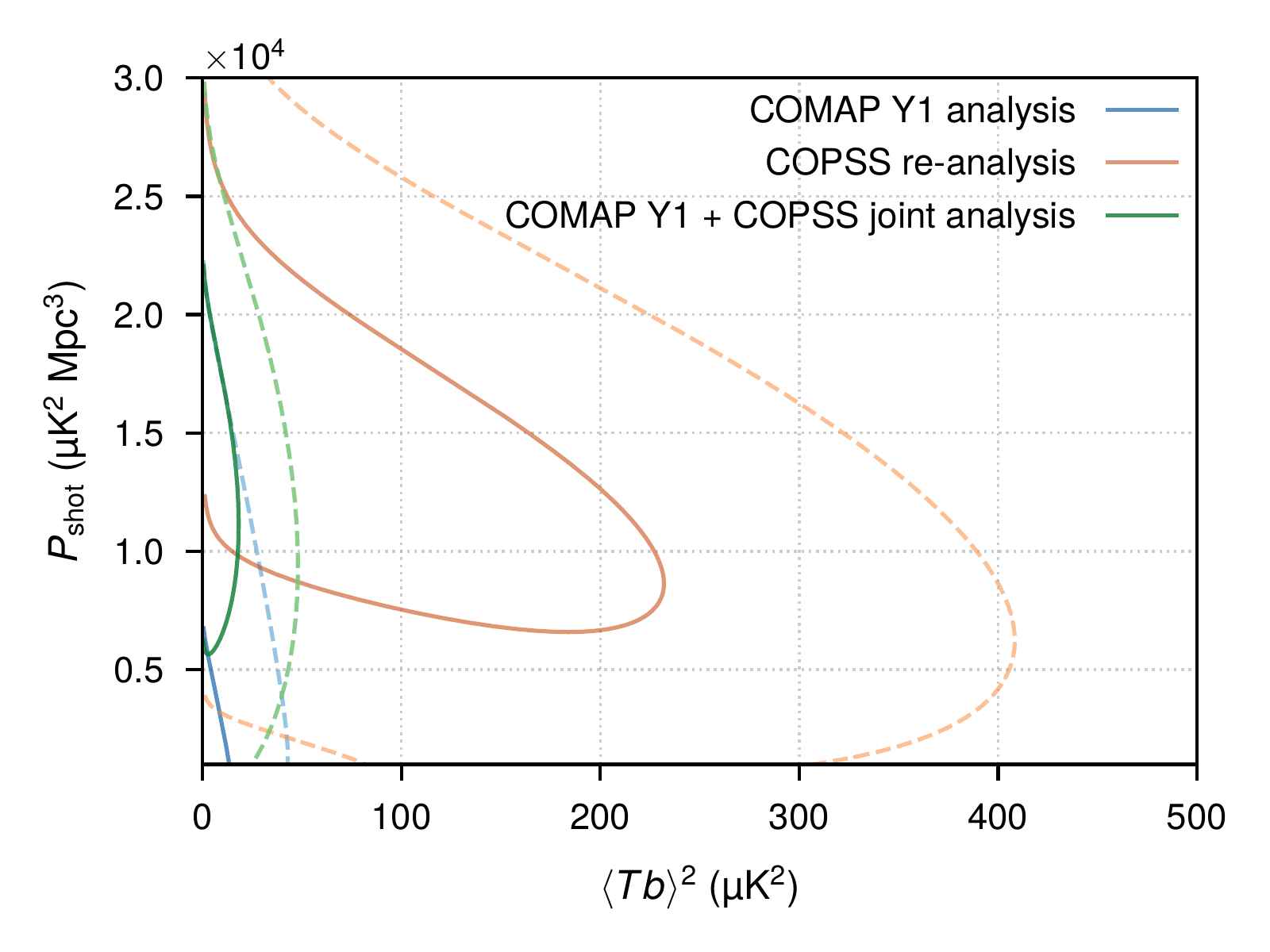}
    \caption{Likelihood contours for the clustering ($\avg{Tb}^2$) and shot-noise amplitudes of the CO power spectrum, based on different datasets. The contours represent $\Delta\chi^2=\{1,4\}$ relative to the minimum $\chi^2$ obtained in the parameter space, representing $1\sigma$ (solid) and $2\sigma$ (dashed) for 2D Gaussians. With the COMAP Season 1 data alone, we obtain an order of magnitude improvement in the constraint on the clustering amplitude.}
    \label{fig:Tbshot_informed}
\end{figure}

To calculate $P(k)$, we obtain average values of $b$ and $v_{\rm eff}$ (an effective velocity scale used to characterize the effect of line broadening; \citealt{chung_etal_21}), from one of our MCMC distributions (``UM + COLDz'') and use them to calculate the effect of redshift-space distortions and line broadening. In this way, we obtain the constraints based on COMAP alone shown in ~\autoref{fig:Tbshot_informed}: $\avg{Tb}^2<49$\,$\mu$K$^2$ and  $P_{\text{shot}} < 24\times10^3\,\mu$K$^2\,$Mpc$^3$. Since COMAP is predominantly sensitive to clustering scales, we do not obtain a very tight constraint on the shot noise component. However, folding in the COPSS result, which is mostly sensitive to shot noise, we obtain only a slightly weaker clustering constraint of $\avg{Tb}^2<51$\,$\mu$K$^2$ and a shot noise constraint of $P_{\text{shot}}  = 11.9^{+6.8}_{-6.1}\times10^3\,\mu$K$^2\,$Mpc$^3$. This estimate is much higher than the shot-noise value from the COPSS re-analysis of $5.8^{+3.2}_{-3.5}\times10^3\,\mu$K$^2\,$Mpc$^3$, partly due to COMAP clustering constraints bounding the shot-noise contribution to COPSS data from below and partly due to corrections for line broadening, which we estimate to attenuate the shot noise measured by COPSS by $\approx40\%$. 

\begin{figure}
    \centering
    \includegraphics[width=0.96\linewidth]{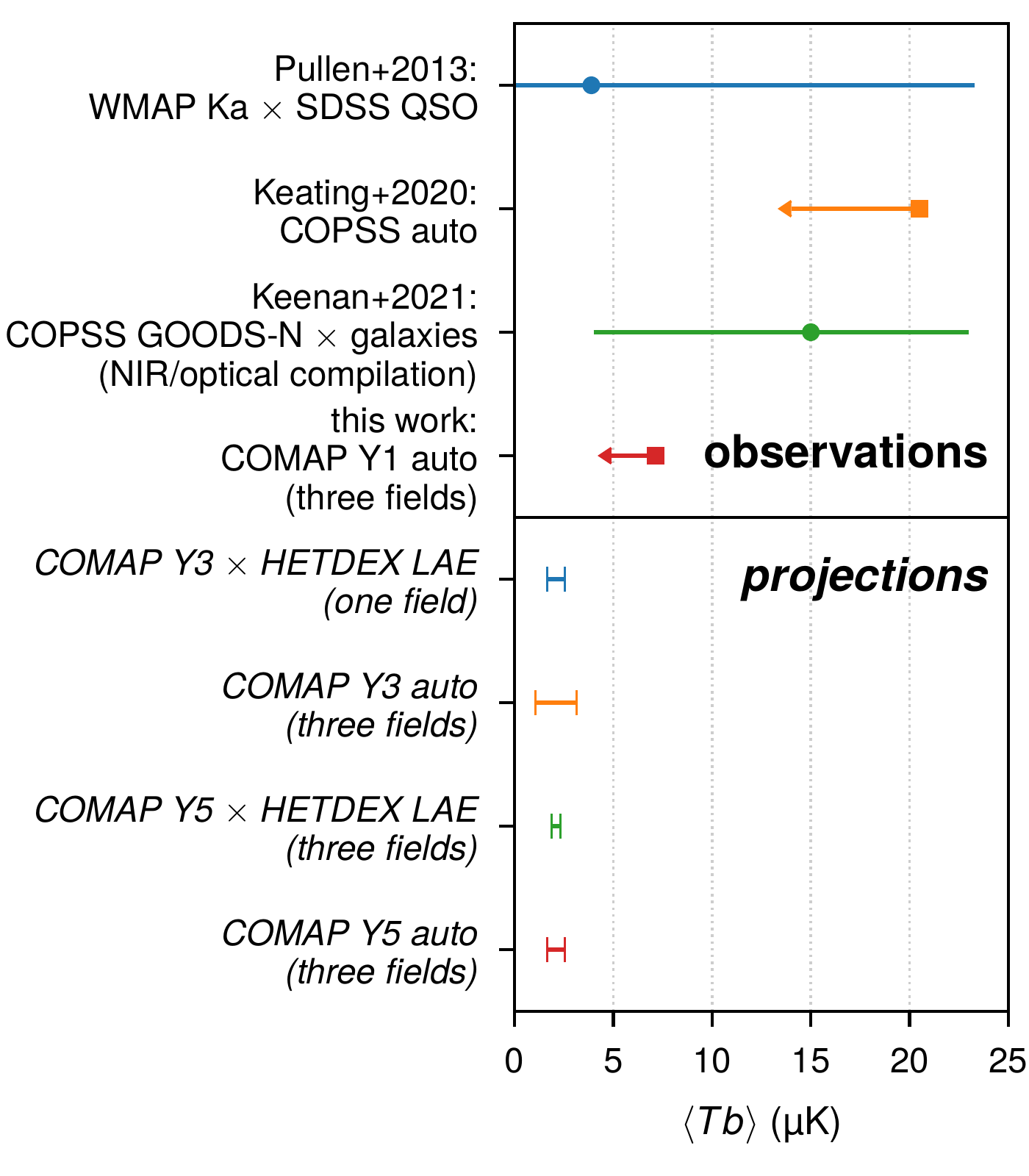}
    \caption{$\avg{Tb}$ constraints from previous observational analyses (upper panel) --- the broadband cross-correlation of~\cite{Pullen13}, the auto-power spectrum of~\cite{keating_etal_16}, and the 3D cross-correlation of~\cite{Keenan21} --- alongside our current upper limit. We also show expected future COMAP constraints (lower panel) for both the CO auto-spectrum and CO--galaxy cross-spectrum based on Fisher forecasts described in \cite{es_V}.}
    \label{fig:Tbconstraints}
\end{figure}

We can obtain a constraint on the H$_2$ mass density by first adopting a value for $b$ and finding the mean CO temperature $\avg{T}$. Conservatively adopting $b>2$ (which is the case for almost all of the sampled parameter sets for all of our priors and for most models in the literature) and combining this constraint with our limit of $\avg{Tb}^2<51$\,$\mu$K$^2$, we find that $\avg{T}<3.6$\,$\mu$K. This is the current best LIM-based constraint on CO(1--0) $\avg{T}$ at $z\sim3$; it represents over a factor of two improvement compared to $\avg{T}<8.2$\,$\mu$K derived from the COPSS CO(1--0) auto-spectrum of \cite{keating_etal_20} and a factor of three improvement compared with $\avg{T}<10.9$\,$\mu$K from the joint COPSS auto-and COPSS--galaxy cross-spectra analysis of from~\cite{Keenan21}. Since the real-space spectrum (before redshift-space distortions) constrains $\avg{Tb}$, it is a useful point of comparison between different analyses that may not necessarily take redshift-space distortions into account. In Figure~\ref{fig:Tbconstraints} we compare the existing constraints on $\avg{Tb}$ and show forecasts for the Pathfinder, including those from a cross-correlation with the HETDEX galaxy survey (to be discussed in \S~\ref{sec:future}).

\begin{figure}[t!]
    \centering
    \includegraphics[width=1.0\linewidth]{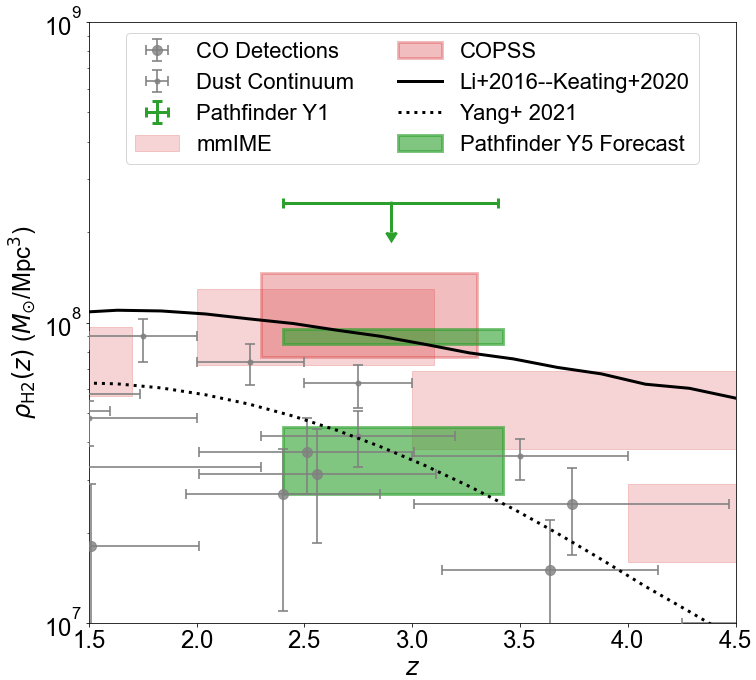}
    \caption{Observational constraints on, and models of, the cosmic molecular gas density in the redshift range $z=2$--$4$. The current COMAP upper limit, and predicted 5-year constraints (green) for two models (Li+2016--Keating+2020 and UM+COPSS+COLDz) representing the range of possible CO signal, are shown compared with those from galaxy surveys (gray) and intensity mapping (red) measurements. The molecular gas history inferred from the Li+2016--Keating+2020 model is plotted as a solid line. Since the UM+COPSS+COLDz model does not include a redshift dependence, we show instead (dotted line) the model of \cite{yang_etal_21} (Yang21) which gives a similar result at $z\sim3$. In both cases we assume a constant $\alpha_{\rm{CO}}=3.6\ M_{\odot}\ (\rm{K}\ \rm{km}\,\rm{s}^{-1}\,{\rm pc}^2)^{-1}$. The galaxy survey points include those compiled by \citet{walter_etal_20} as well as direct CO observations from ASPECS \citep{ASPECS-LPLF2}, COLDz \citep{riechers_etal_19}, and PHIBBS2 \citep{PHIBBS2Lenkic}.  The CO intensity mapping constraints are from the COPSS (dark red) and mmIME (light red) surveys.}
    \label{fig:rhoH2UL}
\end{figure}
Our upper limit on $\avg{T}$ can be expressed as a constraint on the H$_2$ mass density using
\begin{equation}
    \rho_{{\rm H}_2} = \frac{\alpha_{\text{CO}}\avg{T}H(z)}{(1+z)^2},
    \label{eqn:co_h2}
\end{equation}
where $\alpha_{\rm CO}$ converts between H$_2$ mass in units of $M_\odot$ and CO luminosity in units of K\,km\,s$^{-1}$\,pc$^2$, and $H(z)$ is the Hubble parameter at redshift $z$. At the central Pathfinder redshift of $z\approx2.8$, we obtain $\rho_{\text{H}_2}<2.5\times10^8\,M_\odot$\,Mpc$^{-3}$, given $\alpha_{\text{CO}}=3.6\,M_\odot\,($K\,km\,s$^{-1}$\,pc$^2)^{-1}$. We show this constraint along with those from other LIM measurements and galaxy surveys in Figure~\ref{fig:rhoH2UL}. Also plotted are two models representing the range of expected molecular gas density.  Unlike constraints from other methods, ours is based on a direct measurement of CO(1--0) at $z\sim3$ within a large cosmic volume (and so with reduced cosmic variance) and does not make assumptions about CO line ratios. With Season 1 data alone, we obtain a result comparable to that obtained using $>1500$ hours integration time on ALMA calibrators \citep{Klitsch19}, demonstrating the power and scalability of CO LIM.
\begin{figure*}[t]
\begin{center}
\includegraphics[width=\textwidth]{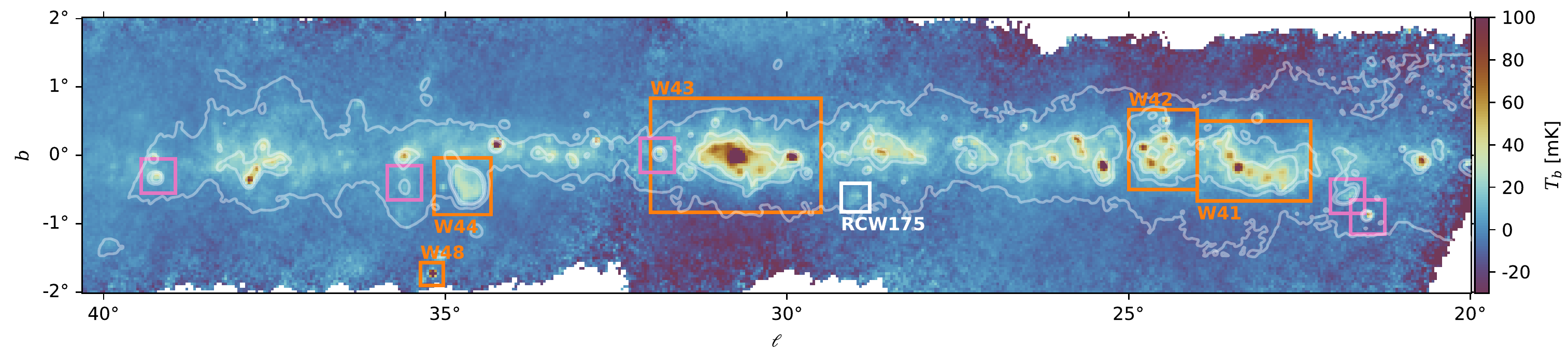}
\caption{The current COMAP band-averaged $30.5$\,GHz map covering the Galactic plane between $20^{\circ}<\ell<40^{\circ}$. The color scale is linear and in units of brightness temperature (mK). Contours are 1.0, 1.5 and 2.0\,MJy/sr from the Parkes 5\,GHz Galactic plane survey \citep{CalabrettaEtAl2014}. Well-known Westerhout star-forming complexes are indicated by orange outlines, including the SNR W44. The other detected SNR are indicated by purple outlines. The RCW175 region, for which a spectral decomposition is shown in Figure~\ref{fig:RCW175}, is indicated with a white outline. Masked pixels are white.}
\label{fig:galmap}
\end{center}
\end{figure*}
\begin{figure}
\begin{center}
\includegraphics[width=0.47\textwidth]{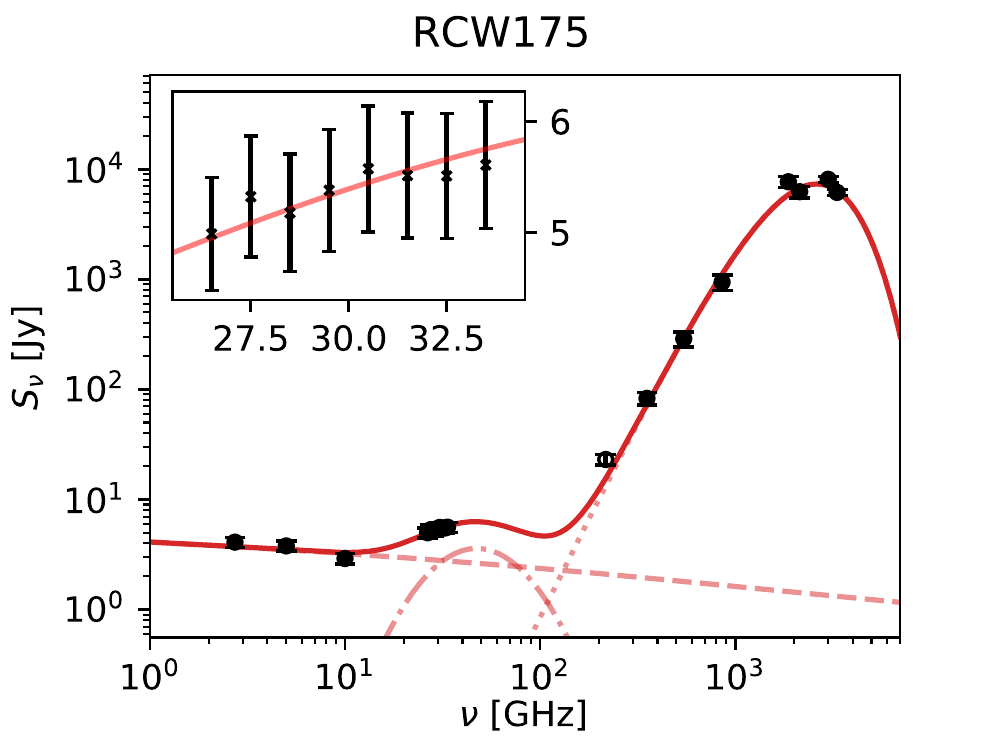}
\caption{Spectral energy distribution for RCW175 featuring COMAP data points between 26--34\,GHz. Using ancillary data covering 2.7\,GHz to 3000\,GHz \citep{es_VI} we have constrained a three-component model consisting of optically thin free-free emission (dashed line), modified blackbody for thermal emission (dotted line) and spinning dust (dot-dashed line). The inset panel shows the COMAP band in more detail. The spinning dust component is detected at $4.9\,\sigma$.}
\label{fig:RCW175}
\end{center}
\end{figure}
\section{First results from the COMAP Galactic Plane Survey} \label{sec:gps}

Although designed to target LIM measurements of redshifted CO, the Pathfinder instrument also brings new capabilities to bear on the study of line and continuum emission from our own and nearby galaxies. At frequencies around 30\,GHz, no other instrument combines the Pathfinder's spatial and spectral resolution with its sensitivity to large angular scales. Observations using the Pathfinder thus fill a gap in existing coverage and have the potential to elucidate the processes of star formation and emission mechanisms of the interstellar medium (ISM), in particular the anomalous microwave emission (AME) which is believed to have its origin in spinning dust grains \citep{leitch_etal_97, dickinson_etal_18}.

During Season 1, around 19\,hrs per day were spent observing the three LIM science fields and calibrators while the remaining time was available for other targets. We used this time to observe individual compact regions of the Galaxy known to exhibit AME as well as other targets of interest, such as M31. These will be the subject of other papers; here, we present initial results from a survey of the Galactic plane (described in more detail in \citealt{es_VI}).

These results are based on observations using the COMAP Pathfinder only during the period 2019 June to 2021 April, covering Galactic longitudes $20^{\circ} < \ell < 40^{\circ}$. Since the end of that period, we have continued to extend the survey along the  Galactic plane and the resulting maps (expected to cover $20^{\circ} < \ell < 220^{\circ}$) will be the subject of future publications.

Because the observing times for the survey were determined by gaps in availability of the three LIM science fields, it was not possible to observe the survey fields during both rising and setting. For this reason, we found that Lissajous scans \citep{es_III} provided the best cross-linking for map-making. As for the LIM pipeline, data analysis for the Galactic plane survey starts from the raw data as recorded by the instrument, together with pointing information and house-keeping data. After this point, an independent pipeline is used since the LIM pipeline removes any continuum signal. 

Initial data selection rejects scans affected by poor weather, identified by examining the feed-feed noise correlation and tracking the $1/f$ noise properties of the data. A running median filter is applied to the TOD to suppress large-scale $1/f$ fluctuations. Ground pickup is removed by fitting a linear slope in azimuth. Calibration is achieved in two steps. Firstly, a relative calibration is performed using the calibration vane; this procedure corrects for atmospheric absorption along the line of sight. Secondly, absolute calibration to the main beam brightness scale is accomplished using observations of the Crab Nebula (Tau-A) and comparing them to spectral fits from the Wilkinson Microwave Anisotropy Probe (WMAP).

A destriping map-maker is used to suppress any residual large-scale contamination in the data. This is accomplished by fitting linear offsets to the TOD and making use of scans in different sky orientations to distinguish between the desired sky signal and $1/f$ noise \citep{Delabrouille1998,Sutton2009,Sutton2010}. This step reduces the $1/f$ noise by at least a factor of four on scales up to 30 arcmin.

Figure~\ref{fig:galmap} shows the COMAP 30.5\,GHz Galactic plane map, covering $20\degr <\ell< 40\degr$. Several bright giant molecular clouds, H\textsc{ii} regions and supernova remnants are detected with high signal-to-noise ratio (S/N). At our observing frequencies, the diffuse emission is dominated by (typically optically thin) free-free emission but there is a significant contribution (20--40\%) from AME \citep{PlanckEarlyXXI,PlanckIntXXIII}. 

By combining the COMAP data with additional surveys at other frequencies, we can decompose the spectral energy distribution (SED) of high S/N regions into contributions from these two emission mechanisms. We interpret any significant excess over that produced from these two components alone as AME and fit it with a simple lognormal curve that approximates more complete parameterizations \citep{Stevenson2014}.

Using this method, we find six regions within our current map that exhibit significant AME. These are discussed in detail in \citet{es_VI}, but we show an example of one such SED in Figure~\ref{fig:RCW175}, for an area containing the bright H\textsc{ii} region RCW175 \citep{Rodgers1960} which has previously been observed to contain AME \citep[see][]{TibbsEtAl2012,BattistelliEtAl2015}. We find that in this region, approximately 50\% of the total flux density at 30\,GHz is due to AME, consistent with other studies \citep{Dickinson2009}.

We also detect six out of the 33 known supernova remnants (SNR) within the currently-surveyed area. Our 1-GHz spectral binning of COMAP data allows us to detect evidence of the steeper spectral indices suggesting spectral aging in these sources. Making use of the native spectral resolution of our data, we have also been able to extract five hydrogen radio recombination lines (RRL) at 20\,km\,${\rm s}^{-1}$ velocity resolution. The resulting RRL survey represents the highest frequency Galactic RRL survey to-date.
\section{Future plans for COMAP} \label{sec:future}

This paper has presented the results from the first season of observing with the Pathfinder (approximately one year). We are continuing to observe and expect to achieve a detection of the clustering component of the CO auto-power spectrum at $z\sim3$ within a total of 5 years of observations and a detection of the CO--galaxy cross-spectrum within a total of 3 years. We briefly present the forecasts leading to this expectation below (see \citealt{es_V} for further details).

The CO signal from the end of the EoR also contributes to our Pathfinder data and we expect to be able to place an interesting limit on this through the use of an overlapping Ly-$\alpha$ galaxy survey data to mask emission from $z\sim3$ in the stacked data \citep{silva_etal_21}. Our ultimate aim, however, is to detect the CO power spectrum from the Epoch of Galaxy Assembly back to the EoR, thereby tracing the global properties of galaxies over cosmic time. The level of the CO signal at such distant epochs is highly uncertain and the subject of continuing investigation. Nevertheless, we examine what current models predict for our prospects to detect such a signal with future phases of the experiment below (see \citealt{es_VII} for further details).

\subsection{Forecasts for the Pathfinder\label{sec:zeq3_forecasts}}

Our observations thus far have enabled us to develop and demonstrate the understanding and control of systematic errors to the level needed to detect the aggregate CO signal from $z\sim3$. En route, we have identified and fixed a number of hardware issues (see \citealt{es_II}). We have also identified a number of remaining issues that can be resolved in the near future, improving the observing efficiency and overall performance (see \citealt{es_III} for details). Taking these into account and considering the expected performance of the system at the end of five years of observing, we expect a factor of $\sim69$ improvement in sensitivity to the auto-power spectrum compared to that obtained in Season 1. Figure~\ref{fig:Pk_mods} shows the predicted sensitivity compared with the fiducial and other models from the literature.

Even for our conservative fiducial model (which likely under-predicts the contribution from faint sources) we expect a detection of the CO(1--0) auto-power spectrum across all $k$ with S/N of 9 at the end of year 5. This rises to S/N of 17 if we consider the Li+2016--Keating+2020 model, which predicts a greater contribution from faint sources. For the fiducial model, Figure~\ref{fig:Tbconstraints} (lower panel) shows the projected constraints on $\avg{Tb}$. After 3 years of observing with the Pathfinder, we expect a marginal $2\sigma$ null rejection of $\avg{Tb}=2.1\pm1.0\,\mu$K, but this will improve to a $\approx5\sigma$ constraint after 5 years. For the Li+2016--Keating+2020 model, the higher expected $P(k)$ S/N would yield correspondingly tighter constraints on $\avg{Tb}$, with significances of $6\sigma$ after 3 years and $12\sigma$ after 5 years. We obtain similar constraints on the molecular gas density, $\rho_{\text{H}_2}$, predicted from these two models (Figure~\ref{fig:rhoH2UL}).

In addition to their potential to detect the CO(1--0) auto-power spectrum, the Pathfinder observations were designed to overlap with the volume covered by HETDEX, a spectroscopic galaxy survey, allowing a CO-galaxy cross-correlation. \cite{chung_etal_19} found that the expected S/N for this cross-spectrum would be significantly greater than that expected for the CO auto-spectrum. In an updated forecast based on the UM+COLDz+COPSS model, \cite{es_V} find that, based on very conservative assumptions, we expect to detect the cross-spectrum over all $k$ with a S/N of 7 after 3 years using Field 1 data alone, rising to 19 after 5 years using data across all three fields. In terms of a constraint on $\avg{Tb}$, this corresponds to a $5\sigma$ detection after 3 years (one field), rising to $10\sigma$ at the end of the 5-year period (three fields; see Figure~\ref{fig:Tbconstraints}). Again, these constraints would be significantly tighter for the Li+2016--Keating+2020 model, at $8\sigma$ and $13\sigma$ after three years (one field) and five years (three fields) respectively.

We note that the large volume covered by COMAP enables multiple voxel-level analyses that will allow us to explore the CO signal in more detail.  \citet{ihle_etal_18} showed that we can use the one-point Voxel Intensity Distribution \citep[VID;][]{breysse_etal_17} to improve our measurement of the CO luminosity function beyond what is possible with the power spectrum alone \citep[see also discussion in][]{es_V}.  The availability of the overlapping HETDEX catalog enables several other analyses \citep{silva_etal_21,breysse_etal_19}. Stacking COMAP voxels containing HETDEX galaxies is expected to result in a detection of the total CO signal with S/N 2--3 times higher than that obtained using all COMAP voxels. New ``conditional VID" measurements examining the intensity distribution within the stacked voxels can be used to further break down the CO luminosity function and how it changes in HETDEX sources. A constraint on the CO(2--1) emission from $z\sim6$ may be obtained by using the HETDEX catalog to mask COMAP voxels containing signal from $z\sim3$. With the high S/N expected for the total stacked CO signal at $z\sim3$, it will be possible to examine the variation of the CO signal and the CO--galaxy cross-spectrum with spatial scale and their evolution with redshift at the peak of cosmic star formation.

\begin{table}
\centering
\caption{\comapeor\ instrument parameters.}
\label{tab:comapeor}
\begin{tabular}{lc}
\hline
\hline
Parameter & Value \\
\hline
Primary reflector diameter & 18 m \\
No.\ of feeds & 19 \\
Polarization & Dual \\
Beam FWHM @ 16\,GHz & 3.7 arcmin \\
System temperature & 20--27\,K \\
Frequency resolution (native) & $\sim2$\,MHz \\
\uline{Frequency bands} & \uline{redshift coverage}\\
Band 1: 12--13\,GHz & 7.8--8.6 \\
Band 2: 13--15\,GHz & 6.7--7.8\\
Band 3: 15--17\,GHz & 5.8--6.7\\
Band 4: 17--20\,GHz & 4.8--5.8\\
\hline
\end{tabular}
\end{table}
\newpage
\subsection{Prospects for \comapeor\label{sec:zeq6_forecasts}}

In order to target the signal from the EoR, we envisage an enhanced instrument, {\comapeor}, consisting of the existing 30\,GHz Pathfinder plus two duplicate receivers mounted on two more 10.4-m telescopes, as well as a 19-feed 16\,GHz receiver mounted on a prototype ngVLA 18-m dish. The 10.4-m dishes are already available for this use at OVRO and the ngVLA dish will be dedicated to COMAP in 2026. The main features of the 30\,GHz instruments are shown in Table~\ref{tab:instrument} and those of the 16\,GHz receiver are listed in Table~\ref{tab:comapeor}. The extra 30\,GHz systems will provide more sensitivity to CO(2--1) at $z=6$--$8$ and CO(1--0) at $z=2.4$--$3.4$, while the 16\,GHz system will open a new window on CO(1--0) at $z=4.8$--$8.6$ (see Figure~\ref{fig:colines}).

\begin{figure*}
\centering
\includegraphics[width=\textwidth]{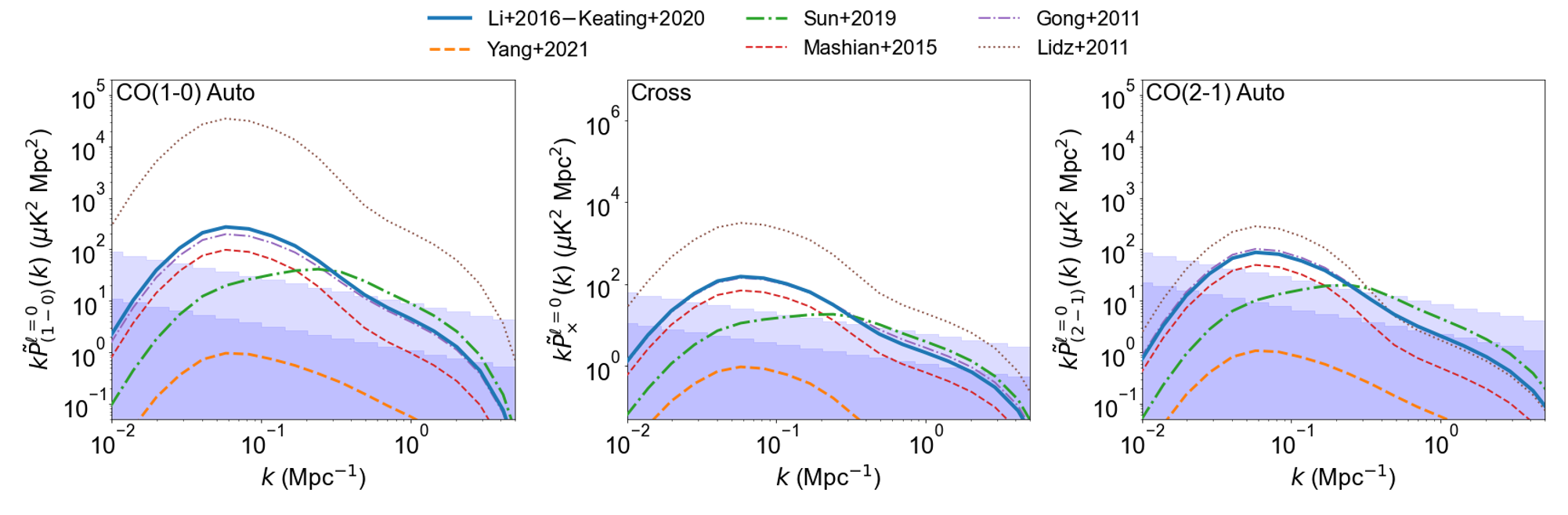}
\caption{Forecast 1-$\sigma$ \comapeor\ and \comapera\ sensitivities (light and dark shaded bands respectively) at $z=6.2$ for models from \citet[][brown dotted]{lidz_etal_11}, \citet[][purple dot-dashed]{gong_etal_12}, \citet[][red dashed]{mashian_etal_15}, \citet[][green dot-dashed]{sun_etal_19}, as well as the Li+2016--Keating+2020 and Yang21 models from Figure~\ref{fig:rhoH2UL}. Predictions are shown for the CO(1--0) auto-spectra (left), the CO(1--0)--CO(2--1) cross-spectra (center), and the CO(2--1) auto-spectra (right).}
\label{fig:pksome}
\end{figure*}

For the purposes of these forecasts, we assume that the 16\,GHz receiver will begin observing at the conclusion of the five-year Pathfinder survey and the two new 30\,GHz receivers two years after that. After five years of observations with all four instruments, we will have obtained twelve years on the Pathfinder, seven years with the 16\,GHz receiver, and five years with each of the new 30\,GHz receivers. This will provide 29,000 dish-hours/field at 30\,GHz (accounting for the $T_{\rm sys}$ adjustment) and 7,000 dish-hours/field at 16\,GHz. We also consider an ``Expanded Reionization Array'' (ERA) concept where the number of dishes is increased to a total of ten at both frequencies, observing for a further five years to produce 110,000 dish-hours  at 30\,GHz  and 57,000 dish-hours at 16\,GHz.

\begin{deluxetable}{cccccc}
\label{tab:SNRb}
\tablecaption{Predicted S/N obtained by \comapeor\ and \comapera\ in each of the instruments' bands (as defined in Table~\ref{tab:comapeor}), along with the combined total (assuming the four frequency bins are independent).  Each entry shows the S/N for both \comapeor\ and \comapera, separated by a slash. The column for Band 3 corresponds to the redshift range for the model predictions of Figure~\ref{fig:pksome}.}
\tablehead{
 & \colhead{Band 1} & \colhead{2} & \colhead{3} & \colhead{4} & \\
\colhead{Redshift} & $7.8$--$8.6$ & $6.7$--$7.8$ & $5.8$--$6.7$ & $4.8$--$5.8$ & \colhead{Total}
}
\startdata
Li16/Keating20 & 2.2/13 & 10/40 & 17/62 & 21/89 & 29/116 \\
Yang21 & 0.0/0.0 & 0.0/0.1 & 0.2/1.1 & 0.6/4.3 & 0.6/4.4 \\
Sun19 & 0.2/1.3 & 2.8/13 & 12/51 & 22/98 & 25/111 \\
Mashian15 & 0.2/1.2 & 2.4/12 & 9.0/32 & 16/60 & 18/70 \\
Gong11 & 0.3/2.4 & 4.4/22 & 16/59 & 30/115 & 34/131 \\
Lidz11 & 52/116 & 114/341 & 179/457 & 356/853 & 418/1030 \\
\enddata
\end{deluxetable}

Figure~\ref{fig:pksome} shows the predicted sensitivity at $z=6.2$ for \comapeor\ and \comapera\ with respect to a number of models for the power spectrum monopole from the literature, including Li+2016--Keating+2020 (which was also adopted for our $z\sim3$ forecasts in \S~\ref{sec:zeq3_forecasts}). Overall, the models span four orders of magnitude in the amplitude of the power spectrum, highlighting our deep ignorance of the processes giving rise to the emission at this redshift and, consequently, the amount that can be learned through observations of this type. 

Table~\ref{tab:SNRb} shows the signal-to-noise ratios expected in each of the four \comapeor\ bands as well as the total (combining detection significances from the monopole, quadrupole and hexadecapole measurements). For all models except the most pessimistic (Yang21), we detect the signal from the EoR at high significance for the \comapeor\ survey and can also resolve the redshift evolution of the signal over the full $z=4.8$--$8.6$ range. \comapera\ produces even higher detection significances for the same three fields although in practice our observing strategy would likely alter to cover wider areas for cross-correlation with other LIM experiments including those targeting \Hi.

Assuming a linear relationship between the CO emission and molecular gas abundance (i.e.\ Equation~\ref{eqn:co_h2}), we can forecast the constraints imposed on the latter by our measurement of the CO power spectrum. As well as the Li+2016--Keating+2020 model, we also consider the most pessimistic, Yang21. The former is consistent with constraints from LIM measurements while the latter is even more pessimistic than constraints from galaxy surveys would suggest.

\begin{figure*}
\centering
\includegraphics[width=0.8\textwidth]{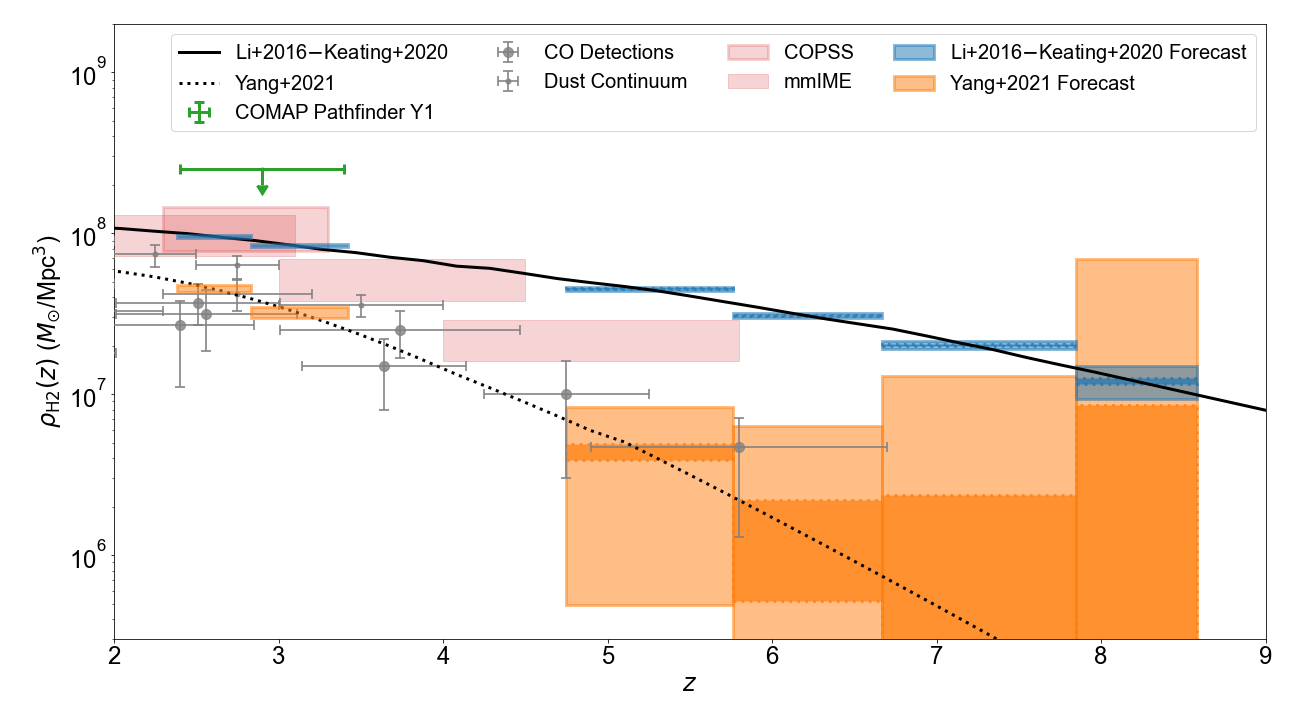}
\caption{Predicted COMAP constraints on the cosmic molecular gas history compared with existing direct and intensity mapping measurements at redshifts covered by future phases of COMAP. The plotted models, data and constraints are as in Figure~\ref{fig:rhoH2UL}. Blue and orange boxes show the 95\% constraints obtained on these models using \comapeor\ (light) and \comapera\ (dark). For reference, we also show the COMAP Y1 constraint from Figure~\ref{fig:rhoH2UL}.}
\label{fig:rhoH2}
\end{figure*}

These models and the predicted constraints for \comapeor\ and \comapera\ are shown in Figure~\ref{fig:rhoH2}, along with existing constraints from galaxy surveys and LIM measurements. \comapeor\ allows us to place very tight constraints on the cosmic molecular gas density in the case where, as suggested by existing LIM measurements, there is a significant contribution from faint galaxies that are missed by current galaxy surveys. For the most pessimistic model, \comapera\ is needed to constrain the molecular gas abundance at high redshift. 

\begin{figure*}
\centering
\includegraphics[width=0.9\textwidth]{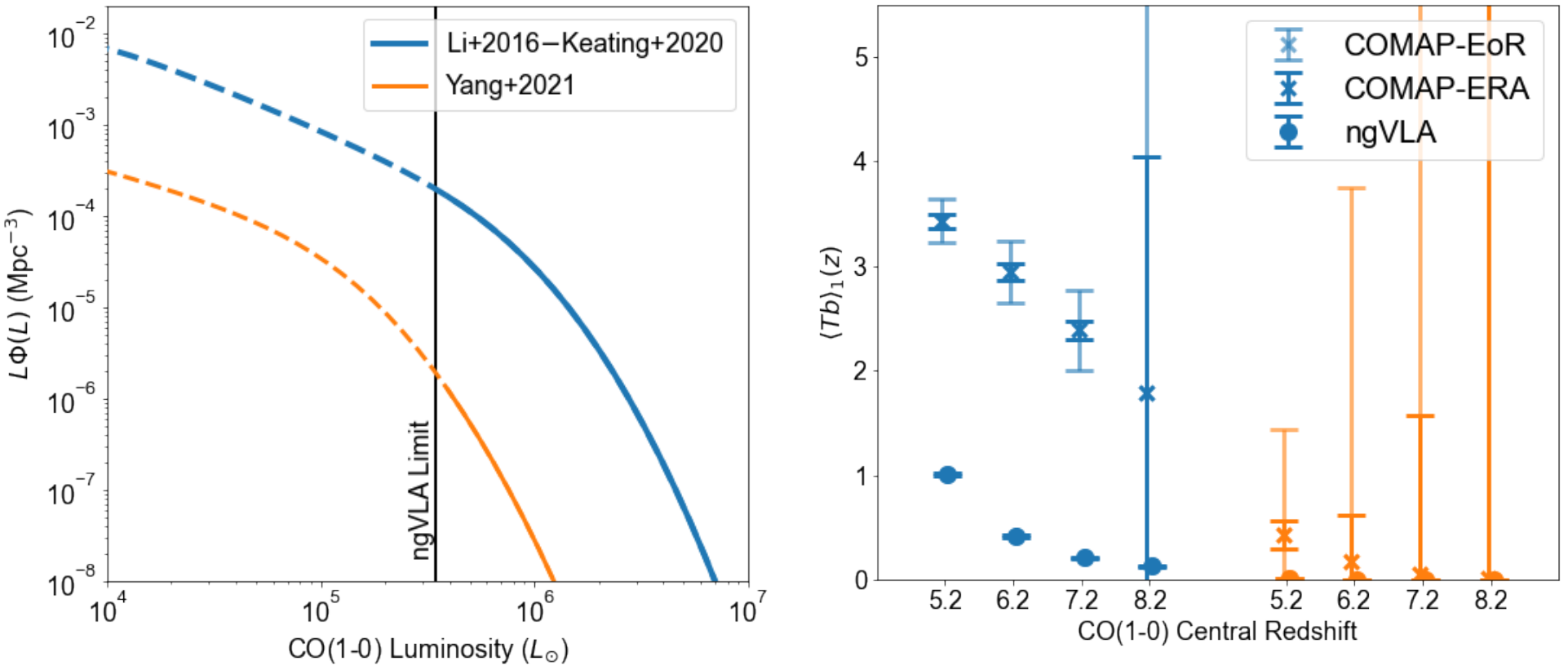}
\caption{{\bf\em Left: }CO(1--0) luminosity functions at $z=6.2$ of the Li+2016--Keating+2020 (blue) and Yang21 (orange) models, with the limit of a hypothetical ngVLA molecular gas survey marked in black.  Dashed lines show the portions of the luminosity functions that are directly accessible only to a LIM survey. {\bf\em Right: }Uncertainties on the CO(1--0) power spectrum amplitude factor for the same two models. Circles show measurements that would be obtained by an ngVLA-like survey, X's show our forecasts for \comapeor\ (light) and \comapera\ (dark).}
\label{fig:FaintGal}
\end{figure*}

The ability of LIM to distinguish between these two cases is a key driver for these observations. Figure~\ref{fig:FaintGal} (left) shows the luminosity functions for our two demonstration models and the detection limit of a hypothetical ngVLA CO(1--0) survey of a deep field as described in \citet{decarli_etal_18}. Galaxies below this limit can be directly detected in aggregate by a LIM measurement. The clustering amplitude inferred from the ngVLA survey (for each model) is shown in Figure~\ref{fig:FaintGal} (left) compared with projected constraints from \comapeor\ and \comapera. For the Li+2016--Keating+2020 model, the value derived from the ngVLA survey under-estimates the clustering amplitude by a large factor. Even if the true high-redshift signal resembles the most pessimistic model (Yang21), LIM observations like COMAP are still required to constrain the possible contribution from faint galaxies.

\newpage
\section{Conclusions\label{conclusions}}

The field of line intensity mapping holds the promise of characterizing the global properties of galaxies over huge spatial volumes across cosmic time. CO has many advantages as a tracer of galaxies for this purpose and provides a measurement that is complementary to galaxy surveys and redshifted 21-cm. A necessary first step on this path is the validation of the technologies as well as the observational and analytical strategies required to detect the CO signal at intermediate redshifts, which has now been accomplished.

We have constructed and fielded a CO LIM Pathfinder and have begun a 5-year survey of redshifted CO(1--0) emission from galaxies at $z$=2.4--3.4. Using data from our first $\sim 13$ months, we have demonstrated the ability of the instrument and our first-generation  analysis pipeline to produce data with the required low levels of correlated noise. 

From our Season~1 data, we obtain a 95\% upper limit of $P_\mathrm{CO}(k) = -2.7 \pm 1.7 \times 10^4\mu\textrm{K}^2\,\mathrm{Mpc}^3$ on scales $k=0.051$--$0.62 \,\mathrm{Mpc}^{-1}$. This is the first direct 3D constraint on the clustering component of the CO(1--0) power spectrum.

This constraint on the auto-power spectrum excludes two models from the literature and resolves ambiguities in the interpretation of the COPSS shot-noise measurement. 

We use this measurement to determine a constraint on the amplitude of the clustering power spectrum,  $\avg{Tb}^2<49$\,$\mu$K$^2$ --- nearly an order of magnitude improvement on the previous best constraint from COPSS. In a re-analysis of the COPSS measurement in tandem with COMAP data, we also obtain a shot noise power, $P_{\text{shot}}  = 1.2^{+0.7}_{-0.6}\times10^4\,\mu$K$^2\,$Mpc$^3$, suggesting that the constraint from COPSS alone of $5.8^{+3.2}_{-3.5}\times10^3\,\mu$K$^2\,$Mpc$^3$ was under-estimated due to line broadening as well as less stringent clustering amplitude constraints.

In addition to pursuing its primary CO LIM science goals, the Pathfinder is also conducting a 30\,GHz survey of the Galactic plane, to cover Galactic longitudes $\ell \sim 20^{\circ}$--$200^{\circ}$ and Galactic latitudes $|b|<1.\!^{\circ}5$, with an angular resolution of $4\farcm5$. We have presented preliminary maps of the plane covering 20\degr$<\ell<40$\degr, demonstrating the ability of the Pathfinder to perform continuum mapping as well as its primary science of spectroscopy. These observations fill a gap in spatial and spectral resolution of the Galactic plane at these frequencies. Performing a spectroscopic decomposition of the emission for point sources in our preliminary map, we find evidence of anomalous microwave emission in a number of these sources. We also detect radio recombination lines at high significance, which will provide a useful resource for the community.

Using the performance of the instrument and analysis pipeline during this first season as a guide and taking into account a range of improvements both already implemented and expected, we forecast the expected constraints on the CO auto-spectrum at the end of a nominal 5-year observing campaign. For the auto-power spectrum, we predict a detection with total S/N of 9--17 after 5 years. For the cross-power spectrum with an overlapping galaxy survey, we predict a S/N of 7 after 3 years from just one field, rising to 19 after 5 years coadding data across all three COMAP fields. 

The availability of an overlapping galaxy catalog also permits various stacking analyses involving selection of COMAP voxels based on their galaxy content, resulting in increased S/N on the stacked CO signal by factors of 2--3 compared to that obtained stacking all COMAP voxels. Similarly, masking COMAP voxels based on their $z\sim3$ galaxy content allows us to place a $6\sigma$ limit on the contribution of CO(2--1) from $z\sim6$.

Looking further ahead, we envisage an enhanced experiment, \comapeor, targeting the EoR directly by adding more pixels at 30\,GHz as well as a new 16\,GHz receiver (on a prototype ngVLA antenna due to be dedicated to COMAP in 2026). Current models for the CO signal at $z\sim6$ span four orders of magnitude, indicating the potential for experimental constraints to improve our understanding of galaxies and the ISM at this epoch. We presented a survey design that provides for a detection at high significance for our fiducial $z\sim6$ model. The resulting constraints on molecular gas abundance have the potential to directly detect or rule out a contribution from faint galaxies that would otherwise be missed by galaxy surveys conducted by future facilities such as the ngVLA. 

We also describe a much more powerful experiment, \comapera, with more feeds at 30 and 16\,GHz, capable of making a $100\sigma$ detection of the EoR CO signal. This could enable a cross-correlation of CO from the ionized ISM of galaxies with 21-cm observations of the neutral IGM, to provide constraints on the evolution of ionized regions during reionization \citep{lidz_etal_11}.

\section*{Acknowledgements}
This material is based upon work supported by the National Science Foundation under Grant Nos.\ 1517108, 1517288, 1517598, 1518282 and 1910999, and by the Keck Institute for Space Studies under ``The First Billion Years: A Technical Development Program for Spectral Line Observations''. Parts of the work were carried out at the Jet Propulsion Laboratory, California Institute of Technology, under a contract with the National Aeronautics and Space Administration, and funded through the internal Research and Technology Development program. DTC is supported by a CITA/Dunlap Institute postdoctoral fellowship. The Dunlap Institute is funded through an endowment established by the David Dunlap family and the University of Toronto. CD and SH acknowledge support from an STFC Consolidated Grant (ST/P000649/1). JB, HKE, MKF, HTI, JGSL, MR, NOS, DW, and IKW acknowledge support from the Research Council of Norway through grants 251328 and 274990, and from the European Research Council (ERC) under the Horizon 2020 Research and Innovation Program (Grant agreement No.\ 819478, \textsc{Cosmoglobe}). 
JG acknowledges support from the University of Miami and is grateful to Hugh Medrano for assistance with cryostat design. 
LK was supported by the European Union’s Horizon 2020 research and innovation program under the Marie Skłodowska-Curie grant agreement No.\ 885990. 
J.\ Kim is supported by a Robert A.\ Millikan Fellowship from Caltech. 
At JPL, we are grateful to Mary Soria for assembly work on the amplifier modules and to Jose Velasco, Ezra Long and Jim Bowen for the use of their amplifier test facilities. 
HP acknowledges support from the Swiss National Science Foundation through Ambizione Grant PZ00P2\_179934. PCB is supported by the James Arthur Postdoctoral Fellowship. RR acknowledges support from ANID-FONDECYT grant 1181620. MV acknowledges support from the Kavli Institute for Particle Astrophysics and Cosmology. We thank Isu Ravi for her contributions to the warm electronics and antenna drive characterization.

\bibliography{CO_bib,COMA_bib,CO_extra,early_science}
\bibliographystyle{aasjournal}

\end{document}